\documentclass[journal]{IEEEtran}

\IEEEoverridecommandlockouts

\usepackage{amsmath,amssymb,amsfonts}
\usepackage{cite}
\usepackage{subfigure}
\usepackage{graphicx}
\usepackage{textcomp}
\usepackage{xcolor}
\usepackage{times}
\usepackage{subfigure}         
\usepackage{amssymb,amsmath}
\usepackage{booktabs}

\usepackage{acronym}  

\usepackage{balance}

\usepackage{epstopdf}

\usepackage{bm}   

\usepackage{algorithm} 

\usepackage{algorithmic}

\usepackage{arydshln}

\usepackage{lipsum}
\usepackage{stfloats}
\usepackage{url}

\usepackage{setspace}
\usepackage{hyperref}

\hypersetup{
	colorlinks = true, %
	linkcolor = black, %
	urlcolor = blue,%
	citecolor = blue} %
\allowdisplaybreaks[4]

\newtheorem{lemma}{\bf Lemma}
\newtheorem{corollary}{\bf Corollary}
\newcommand{\mb}[1]{{  \mathbf  #1}}  

\definecolor{BLUE}{rgb}{0,0,1}

\acrodef{siso}[SISO]{single-input single-output}%

\acrodef{nmse}[NMSE]{normalized mean square error}%

\acrodef{ris}[RIS]{reconfigurable intelligent surface}%

\acrodef{mm}[MM]{majorization-minimization}%
\acrodef{IF}[IF]{ice-filling}%

\acrodef{rhs}[RHS]{reconfigurable holographic surface}%

\acrodef{csi}[CSI]{channel state information}%
\acrodef{awgn}[AWGN]{additive white Gaussian noise}%

\acrodef{mim}[MIM]{mutual-information-maximization}%

\acrodef{cs}[CS]{compressed sensing}%
\acrodef{ao}[AO]{alternaing optimization}%

\acrodef{bs}[BS]{base station}%
\acrodef{snr}[SNR]{signal-to-noise ratio}%
\acrodef{mmwave}[mmWave]{millimeter-wave}%
\acrodef{snr}[SNR]{signal-to-noise ratio}%
\acrodef{rf}[RF]{radio frequency}%

\acrodef{das}[DAS]{dense array system}%

\acrodef{omp}[OMP]{orthogonal matching pursuit}%
\acrodef{mp}[MP]{message passing}%

\acrodef{ml}[ML]{maximum likelihood}%

\acrodef{mse}[MSE]{mean square error }%

\acrodef{as}[AS]{antenna selection}%

\acrodef{vamp}[VAMP]{vector approximate message passing}%

\acrodef{fas}[FAS]{fluid antenna system}%

\acrodef{sinr}[SINR]{signal-to-interference-plus-noise ratio}%

\acrodef{mimo}[MIMO]{multiple-input multiple-output}%
\acrodef{miso}[MISO]{multiple-input single-output}%

\makeatletter
\def\firstletterparse#1#2&{\def\strfirstletter{#1}\def\strotherletters{#2}}

\newcommand{\MakeSmallcaps}[1]{%
	\expandafter\firstletterparse#1&
	\expandafter\MakeUppercase\strfirstletter\textsc{\strotherletters}%
}

\makeatother

\def\BibTeX{{\rm B\kern-.05em{\sc i\kern-.025em b}\kern-.08em
		T\kern-.1667em\lower.7ex\hbox{E}\kern-.125emX}}

\begin{document}
	\title{Observation Matrix Design for Densifying MIMO Channel Estimation via 2D Ice Filling
	}
\author{
{Zijian Zhang and Mingyao Cui
}
\vspace{-1em}
\thanks{
The conference version of this paper was presented in part at the IEEE ICC’25, Montreal, Canada, Jun. 2025 \cite{Zijian'ICC'25}.
}
\thanks{Zijian Zhang is with the Department of Electronic Engineering, Tsinghua University, Beijing 100084, China (e-mail: zhangzj20@mails.tsinghua.edu.cn).}
\thanks{Mingyao Cui is with  the Department of Electrical and Electronic 
Engineering, The University of Hong Kong, Hong Kong (e-mail: mycui@eee.hku.hk).
}
}	
\maketitle
\begin{abstract}
In recent years, densifying multiple-input multiple-output (MIMO) has attracted 
much attention from the communication community. Thanks to the subwavelength 
antenna spacing, the strong correlations among densifying antennas provide 
sufficient prior knowledge about channel state information (CSI). This 
inspires the careful design of observation matrices 
(e.g., 
transmit precoders and receive combiners), that exploits the CSI prior 
knowledge, to 
boost channel estimation performance. Aligned with this vision, this work 
proposes to 
jointly design the combiners and precoders by maximizing the mutual 
information between the received pilots and densifying MIMO channels. 
A two-dimensional ice-filling (2DIF) algorithm is proposed to efficiently 
accomplish this objective. 
The algorithm is motivated by the fact that the eigenspace of MIMO 
channel 
covariance can be decoupled into two sub-eigenspaces, which are 
associated with the correlations of transmitter antennas and receiver 
antennas, respectively. 
By properly setting the precoder and the combiner as the eigenvectors from 
these two sub-eigenspaces, the 2DIF promises to generate near-optimal 
observation matrices. 
Moreover, we further extend the 2DIF method to the popular hybrid combining 
systems, where a two-stage 2DIF (TS-2DIF) algorithm is developed to handle the 
analog combining circuits realized by phase shifters. 
Simulation 
results demonstrate that, compared to the state-of-the-art schemes, the 
proposed 2DIF and TS-2DIF methods can achieve superior channel estimation 
accuracy.
\end{abstract}

\section{Introduction}
In recent years, densifying \ac{mimo} has attracted considerable attention 
from the wireless communication community 
\cite{HuangChongWen'20'WCOM,YutongZhang'23,Zijian'23'JSAC,kanbaz2024,zhang2023reconfigurable,wong2021fluid}.
 Different from the conventional \ac{mimo} whose antennas are usually spaced of 
half wavelength $\lambda/2$, the antenna spacing of densifying \ac{mimo} is 
much smaller, such as $\lambda/6$ \cite{Ovejero'17'TAP}, $\lambda/8$ 
\cite{hwang2020binary}, $\lambda/10$ \cite{Christos'18'COMMAG}, or even 
$\lambda/23$ \cite{liu2019deeply}. By densely arranging massive 
subwavelength-spaced antennas in a compact space, densifying \ac{mimo} promises 
to realize the ultimate control of the radiated/received electromagnetic waves 
on limited apertures. To this end, many dense-antenna transceiver architectures 
have emerged, such as holographic MIMO (H-MIMO) \cite{Gong'24HMIMOSurv}, 
holographic reconfigurable surfaces (RHSs) \cite{YutongZhang'23}, 
continuous-aperture MIMO (CAP-MIMO) \cite{Zijian'23'JSAC}, superdirective 
antenna arrays \cite{kanbaz2024}, reconfigurable intelligent surfaces (RISs) 
\cite{Basar'24'RISMag}, and fluid antenna systems (FASs) 
\cite{wong2021fluid}.  
{ Particularly, in high-frequency (e.g., millimeter-wave or terahertz) communications, densifying MIMO systems are common due to their shorter wavelengths, reduced grating lobes, and enhanced beamforming precision \cite{HuangChongWen'20'WCOM}.}
Utilizing the extensive channel observations facilitated by a multitude of antennas, densifying \ac{mimo} is anticipated to achieve significant array gains and multiplexing-diversity gains \cite{WeiLi'JSTSP'22, AnJiancheng'CL'2023'I, AnJiancheng'CL'2023'II, liu2023densifying}. Furthermore, densifying \ac{mimo} can mitigate the effects of grating lobes and offer enhanced performance for large oblique angles of incidence \cite{di2023mimo}. Some studies have also highlighted their capabilities to realize super-directivity \cite{xie2023genetic, kanbaz2024} or super-bandwidth \cite{Heath'23'JSAC} in wireless transmissions.

Enabled by their phase shifters and \ac{rf} chains,
the transmission performance of \ac{mimo} is determined by the constructive 
precoders/combiners at transceivers \cite{Tse2005WC}. To implement effective 
precoding/combining, an indispensable technology for \ac{mimo} systems is the 
acquisition of \ac{csi} \cite{cui2022near,Tianyue2024CBT}. To date, 
numerous technologies have been proposed to estimate the channels of classical 
\ac{mimo} systems. For example, when the available pilot length exceeds the 
number of antennas, some classical estimators \cite{kay1993fundamentals}, such 
as the least square (LS) estimator and the minimum mean square error (MMSE) 
estimator, can be used to recover \ac{mimo} channels in a non-parametric way. 
By exploiting the channel sparsity in the angular domain, compressed sensing 
(CS)-based channel estimators can enhance the estimation accuracy and reduce 
the pilot overhead \cite{Chongwen'19'TSP,Malong'20'TCSP}. Relevant techniques 
include the orthogonal matching pursuit (OMP)-based estimator 
\cite{Cui'22'TCOM} and the 
approximate message passing (AMP)-based estimator 
\cite{rangan2019vector,Chongwen'19'TSP}. Additionally, some deep learning (DL) 
approaches, which involve training neural networks based on channel datasets, 
are utilized to realize data-driven channel estimation in \ac{mimo} systems 
\cite{Jin'19,Ziwei'20'TVT,Xisuo'21'JSAC}.

Although many channel estimators in the literature can be adopted in densifying 
\ac{mimo} systems, they often exhibit a non-negligible performance gap compared 
to the optimal estimator \cite{kay1993fundamentals}. This is because most 
existing estimators overlook the strong correlations among densifying \ac{mimo} 
antennas. Specifically, since the antenna spacing of densifying \ac{mimo} is 
very small, the channels associated with close-by antennas are spatially 
similar \cite{liu2023densifying}. Besides, the circuit mutual coupling induces 
signal interactions between adjacent antennas, further enhancing the channel 
correlation in densifying \ac{mimo} systems 
\cite{Ovejero'17'TAP,hwang2020binary,Christos'18'COMMAG,liu2019deeply}. These 
facts lead to the highly structured covariance matrices of densifying \ac{mimo} 
channels. Existing works have revealed that, such an informative covariance 
matrix can provide appreciable prior knowledge for the specific design of 
observation matrices (e.g., combiners and precoders) in channel estimation, 
thus significantly 
improving the accuracy of \ac{csi} acquisition 
\cite{williams1995gaussian,CBC'FAS'JSAC,AnJiancheng'CL'2023'III}. 

To exploit the strong channel correlations for improved \ac{csi} acquisition, 
our prior work \cite{cui2024nearoptimal} proposes an ice filling 
(IF) based observation matrix design in \acp{das}, which is
inspired by the idea of Gaussian Process Regression (GPR). 
By 
maximizing the mutual information (MI) characterized by the channel covariance matrix, 
the IF algorithm  can sequentially produce the observation vectors of receivers 
in 
a pilot-by-pilot manner. Through optimizing pilot allocation to the channel 
covariance 
eigenvectors, this method works like filling ice blocks onto different 
orthogonal channels. Then, the designed observation matrix is shown to have
near-optimal channel estimation performance in \acp{das}, which achieves 
much higher estimation accuracy compared to the state-of-the-art schemes 
\cite{cui2024nearoptimal}. 
Nevertheless, despite its ability to exploit the channel 
covariance, IF method is only feasible to 
design the \emph{vector-form receive combiner} in single-input multiple-output 
(SIMO) 
system 
with a single-antenna transmitter and a single-\ac{rf}-chain receiver 
\cite{cui2024nearoptimal}. 
For a general densifying \ac{mimo} system with multiple antennas and multiple 
\ac{rf} 
chains at both transceivers, \emph{the receive and transmit channel covariance 
pair}, as well as the \emph{matrix-form receive combiner and transmit  precoder 
pair}  
are coupled together. As the IF scheme fails to tackle these couplings, 
it is far from optimal for desifying MIMO. To the best of our knowledge, the 
full 
exploitation of densifying \ac{mimo}'s channel covariance for designing 
observation 
matrices is still an unaddressed challenge. 

To fill in this gap, this work generalizes the IF scheme to a two-dimensional 
ice filling (2DIF) scheme, whose core idea is 
to design precoders and combiners by decoupling the coupled channel 
covariance matrices in their eigenspace. Our key contributions and findings are 
summarized as follows.
\begin{itemize}
	{\color{black}
	\item {\bf Generalized framework of observation matrix 
	design}: 
	Inspired by the IF algorithm, we apply the technique of GPR into densifying 
	MIMO channel estimation. 
	Our key idea is to maximize the MI between the received 
	pilots and the MIMO channel by jointly optimizing the receive combiners and 
	transmit precoders. The formulated observation matrix design problem is 
	shown to be a generalization of that discussed in 
	IF~\cite{cui2024nearoptimal} because of the 
	consideration of practical MIMO systems with multi-RF-chain receivers. 
	To be specific, the receiver-side channel 
	covariance considered by IF is generalized to a MIMO channel covariance, which relies on the correlations at both sides of the transceivers. 
	Moreover, the observation matrix is no longer a vector-form 
	combiner, 
	but the Kronecker product of the matrix-form combiner and precoder. These properties fundamentally
	distinguish our design from the IF scheme. 

	\item {\bf 2DIF based observation matrix design}: To overcome the design 
	challenges imposed by the coupling of matrix-form combiners and precoders 
	in observation matrices, a 2DIF based observation matrix design is 
	developed. The proposed design employs a greedy method to  
	jointly produce the combiners and precoders in 
	a block-by-block way. Concretely, we first prove that the eigenspace of the 
	channel covariance can be decoupled into two sub-eigenspaces, which are 
	associated with the correlations of transmitter antennas and receiver 
	antennas, respectively. Then, utilizing the eigenspace invariance, we show 
	that the near-optimal observation matrix can be obtained by properly 
	setting the 
	precoder and the combiner as the eigenvectors from these two 
	sub-eigenspaces, which can be realized by a linear search algorithm. 
	Besides, we also provide an intuitive and insightful explanation for 2DIF 
	to clarify its physical significance. Similar to the water-filling 
	precoding which maximizes the \ac{mimo} capacity, the implementation of 
	2DIF can be viewed as a two-dimensional ice-filling process.
}

	\item {\bf TS-2DIF based observation matrix design}: The proposed 2DIF 
	method requires that the amplitude of each receiver antenna can be 
	controlled independently. However, for many hybrid \ac{mimo} structures, 
	only the phase shifts of analog combiners can be reconfigured, thus the 
	proposed 2DIF cannot be directly adopted in these scenarios. To address 
	this issue, we propose the two-stage 2DIF (TS-2DIF) method. 
	Different from the exiting works on hybrid \ac{mimo} beamforming that precoders and combiners can be designed independently, in our considered observation matrix design, the beamformer
	and the combiner are coupled due to a Kronecker-product structure. To overcome this challenge, the proposed TS-2DIF alternately optimizes the analog combiner, the digital 
	combiner, and the precoder at the transceivers to approach the ideal observation matrix.
\end{itemize}

The rest of this paper is organized as follows. In Section \ref{sec:model}, the system model and problem formulation are introduced. In Section \ref{sec:2DIF}, the proposed 2DIF based observation matrix design for channel estimation is illustrated. In Section \ref{sec:TS-2DIF}, the proposed TS-2DIF based observation matrix design is provided. In Section \ref{sec:KS}, the computational complexities of the proposed methods are analyzed, and the kernel selection is discussed. In Section \ref{sec:sim}, simulations are carried out to verify the effectiveness of the proposed schemes. In Section \ref{sec:con}, conclusions are drawn.

\textit{Notation:} ${[\cdot]^{T}}$, ${[\cdot]^{H}}$, ${[\cdot]^{*}}$, and ${[\cdot]^{-1}}$ denote the transpose, conjugate-transpose, conjugate, and inverse operations, respectively; $\|\cdot\|$ denotes the $l_2$-norm operation; $\|\cdot\|_F$ denotes the Frobenius-norm operation; ${z}(i)$ denotes the $i$-th entry of vector ${\bf z}$; ${\bf Z}({i,j})$, ${\bf Z}({j,:})$ and ${\bf Z}({:,j})$ denote the $(i,j)$-th entry, the $j$-th row, and the $j$-th column of matrix ${\bf Z}$, respectively; ${\rm Tr}(\cdot)$ denotes the trace of its argument; ${\rm diag}(\cdot)$ and ${\rm blkdiag}(\cdot)$ are the diagonal and the block-diagonal operations, respectively; ${\mathsf E}\left(\cdot\right)$ is the expectation operator; $\Re\{\cdot\}$ denotes the real part of the argument; $\ln(\cdot)$ denotes the natural logarithm of its argument; $\mathcal{C} \mathcal{N}\!\left({\bm \mu}, {\bf \Sigma } \right)$ denotes the complex Gaussian distribution with mean ${\bm \mu}$ and covariance ${\bf \Sigma }$; ${\cal U}\left(a,b\right)$ denotes the uniform distribution between $a$ and $b$; $\mathbf{I}_{L}$ is an $L\times L$ identity matrix; $\mathbf{1}_{L}$ is an all-one vector or matrix with dimension $L$; and $\mathbf{0}_{L}$ is a zero vector or matrix with dimension $L$.

\section{System Model and Problem Formulation}\label{sec:model}
\begin{figure}[!t]
	\centering
	\includegraphics[width=0.47\textwidth]{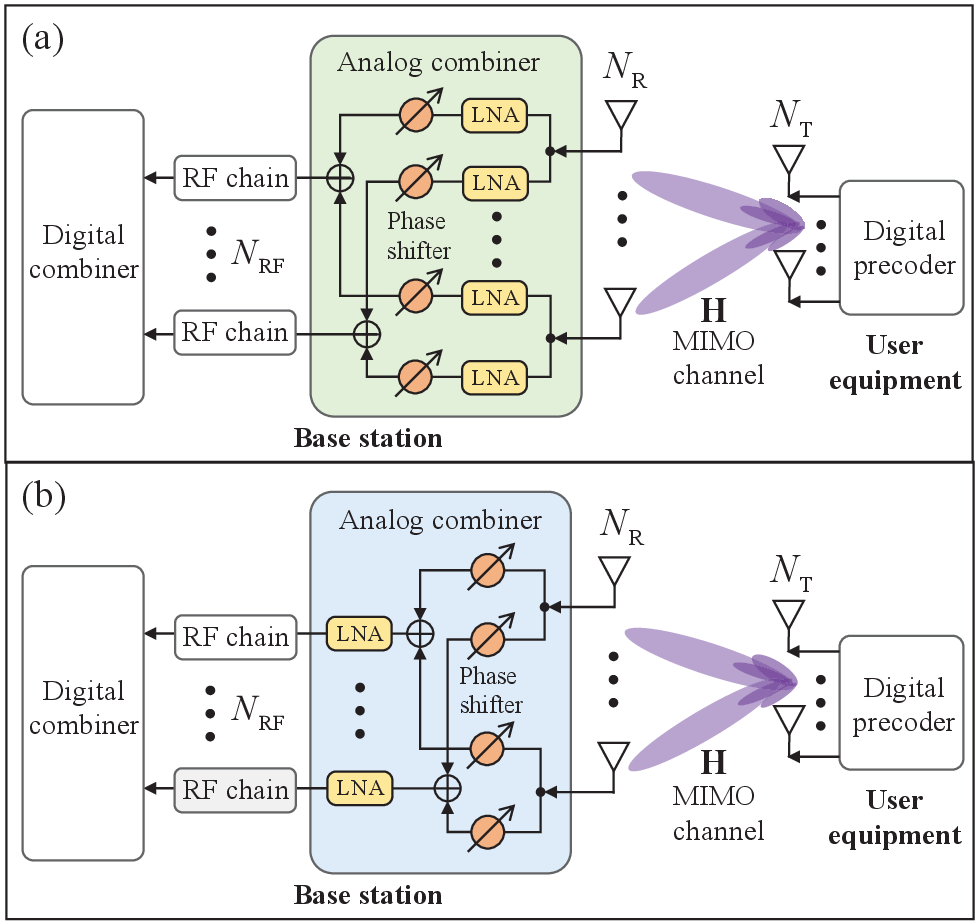}
	\caption{An illustration of hybrid analog and digital \ac{mimo} architectures, where the structure of BS is built on (a) an amplitude-and-phase 
		controllable analog 
		combiner and (b) a phase-only controllable analog combiner, respectively. 
	}
	\label{img:system}
\end{figure}


\subsection{Transceiver Model}
This paper considers the uplink channel estimation of a densifying \ac{mimo} 
system, consisting of an $N_{\rm R}$-antenna \ac{bs} equipped with $N_{\rm RF}$ 
\ac{rf} chains and an
$N_{\rm T}$-antenna user. The antennas at 
transceivers are densely arranged with sub-wavelength antenna spacing $d$. We 
define 
${\bf 
H}\in{\mathbb C}^{N_{\rm R}\times N_{\rm T}}$ 
as the wireless channel and $Q$ as the number of transmit pilots within a 
coherence-time frame. The received
signal ${\bf y}_q\in{\mathbb C}^{N_{\rm RF}}$  at the \ac{bs} in 
timeslot $q$ 
is modeled as 
\begin{align}\label{eqn:y_p}
{\bf y}_q &= {\bf W}^H_q{\bf H}{\bf v}_qs_q + {\bf W}^H_q{\bf z}_q \notag\\&= 
\left( 
{{\bf{v}}_q^T \otimes {\bf{W}}_q^H} \right){\bf{h}}s_q + 
{\bf{W}}_q^H{{\bf{z}}_q}, 
\end{align}
where  ${\bf h} \equiv {\rm vec}({\bf H})$ is defined as the  
vectorized channel matrix, $s_q$  the pilot symbol, and ${\bf z}_q \sim \mathcal{C} 
\mathcal{N}\!\left({\bf 0}_{M},  \sigma^2{\bf I}_{M} \right)$ is the 
\ac{awgn}. 
Vector ${\bf v}_q\in{\mathbb C}^{N_{\rm T}}$ denotes the precoder at the user. 
For the transmitter, the user equipment typically employs a fully digital 
precoder with a moderate number of antennas, $N_T$. 
 Thereby, the coefficient of $\mb{v}_q$ can be freely configured as long 
 as the power constraint 
is satisfied: $\|\mb{v}_q\|^2 = P$, with $P$ the per-pilot transmit power.
For the receiver, ${\bf W}_q:= \mb{A}_q\mb{D}_q \in{\mathbb 
	C}^{N_{\rm R}\times N_{\rm RF}}$ is the hybrid combiner at the \ac{bs}, 
	with $\mb{A}_q \in{\mathbb C}^{N_{\rm R}\times N_{\rm RF}}$ and $\mb{D}_q 
	\in{\mathbb C}^{N_{\rm RF}\times N_{\rm RF}}$ being the analog and digital 
	combiners, respectively. As presented in 
	Fig.~\ref{img:system}, we focus on two typical 
	implementations of the analog combiners: the amplitude-and-phase 
	controllable combiner and the phase-only controllable combiner. The former 
	deploys one phase shifter and one low-noise-amplifier (LNA) between 
	each antenna-to-RF chain link. In this 
	architecture, both the amplitude and phase of the elements of $\mb{A}_q$ 
	are adjustable, 
	and thus the coefficients of $\mb{W}_q$ can be freely controlled. The 
	latter, however, employs one phase shifter to connect each antenna-to-RF chain link, 
	while the signals aggregated on each RF chain are jointly processed by a 
	global LNA.
In this context, only 
	the phases of 
	$\mb{A}_q$ are 
	adjustable, which imposes a structural constraint on the feasible set of 
	$\mb{W}_q = \mb{A}_q\mb{D}_q$. 

Without loss of 
generality, 
 we assume $s_q = 1, 
\forall q\in\{1,\cdots,Q\}$. Considering the total $Q$ timeslots for pilot 
transmission, we arrive at 
\begin{equation}\label{eqn:y}
	{\bf y} = {\bf X}^{H}{\bf h} + {\bf z},
\end{equation}
where ${\bf y} := \left[{ \bf y}_1^T,\cdots,{\bf y}_Q^T\right]^T$,  ${\bf z} := 
\left[ {\bf z}_1^H{\bf 
	W}_1,\cdots, {\bf z}_Q^H{\bf W}_Q \right]^H$, ${\bf{X}} = 
\left[{\bf{X}}_1,\cdots, {\bf{X}}_Q\right]$, and ${\bf X}_q:={{\bf{v}}_q^* 
	\otimes {\bf{W}}_q}$ is defined as the observation matrix for each pilot. 
	This paper aims at accurately estimating $\mb{h}$ from $\mb{y}$ by jointly designing combiners $\{\mb{W}_q\}_{q=1}^Q$ and precoders $\{\mb{v}_q\}_{q=1}^Q$.


\subsection{Channel Model}
We consider the general correlated Rayleigh-fading channel model, which simultaneously takes the spatial correlation, spherical wavefront, and electromagnetic mutual coupling into account. Following this well-known model \cite{wu2008effects,Chen'16'TVT,kolomvakis2024exploiting}, the channel ${\bf H}$ between the BS and the user is expressed as:
\begin{align}\label{eq:channel_model}
{\bf H} = {\bf C}_{\rm rx}^{1/2}{\bf R}_{\rm rx}^{1/2}{\bf H}_{\text{iid}}{\bf R}_{\rm tx}^{1/2}{\bf C}_{\rm tx}^{1/2},
\end{align}
where ${\bf R}_{\rm rx}\in{\mathbb C}^{N_{\rm R}\times N_{\rm R}}$ and ${\bf R}_{\rm tx}\in{\mathbb C}^{N_{\rm T}\times N_{\rm T}}$ are the spatial correlation matrices at the BS and the user, respectively; ${\bf C}_{\rm rx}\in{\mathbb C}^{N_{\rm R}\times N_{\rm R}}$ and ${\bf C}_{\rm tx}\in{\mathbb C}^{N_{\rm T}\times N_{\rm T}}$ characterize the electromagnetic mutual coupling among the BS antennas and the user antennas, respectively; and ${\bf H}_{\text{iid}}$ is a complex Gaussian matrix in which each elements are i.i.d with zero mean and unit variance. This channel model is widely recognized by the studies on correlated MIMO channels, also adopted by commercial channel simulators, such as Matlab Antenna toolbox \cite{MathWorks5GToolbox}.

\textbf{Example 1} (Example of settings).
Note that, the settings of ${\bf R}_{\rm rx}$, ${\bf R}_{\rm tx}$, ${\bf C}_{\rm rx}$, and ${\bf C}_{\rm tx}$ depend on the specific transmission scenarios in practice. As a typical example, the $(m,n)$-th entry of ${\bf R}_{\rm rx}$ is usually modeled as \cite{Gong'24HMIMOSurv,AnJiancheng'CL'2023'I,Pizzo'22'TWC}:
\begin{align}\label{eqn:R_rx_m_n}
	\mathbf{R}_{\rm rx}(m, n)=\int \int_{-\frac{\pi}{2}}^{+\frac{\pi}{2}} f_{\rm rx}(\varphi, \theta) e^{j \mathbf{k}(\varphi, \theta)^T\left(\mathbf{r}_m-\mathbf{r}_n\right)} {\rm d} \theta {\rm d} \varphi,
\end{align}
where $f_{\rm rx}(\varphi, \theta)$ is the scenario-dependent spatial-scattering function, which describes the angular multipath distribution and the directivity gain of receiver antennas in practice; $\mathbf{k}(\varphi, \theta) := \frac{2 \pi}{\lambda}(\cos \theta \cos \varphi, \cos \theta \sin \varphi, \sin \theta)^T$ is the wave vector with $\lambda$ being the wavelength; and ${\bf r}_m\in{\mathbb R}^3$ is the location of the $m$-th receiver antenna. To characterize the mutual coupling, ${\bf C}_{\rm rx}$ is modeled as \cite{wu2008effects,gupta2003effect}:
\begin{align}
\mathbf{C}_{\rm rx}=\left(\mathbf{Z}_{\rm rx}+R_{\rm rx} \mathbf{I}\right)^{-1},
\end{align}
where $\mathbf{Z}_{\rm rx}$ is the mutual impedance matrix and $R_{\rm rx}>0$ is the dissipation resistance of antennas. Given the specific array parameters, $\mathbf{Z}_{\rm rx}$ and $R_{\rm rx}$ can be calculated by the well-known induced electromagnetic fields (EMF) method \cite{balanis2016antenna}. Similar settings can be adopted to generate ${\bf R}_{\rm tx}$ and ${\bf C}_{\rm tx}$.	

\subsection{Problem Formulation}
As antennas of densifying \ac{mimo} are packed within a small area, the 
channels across close-by antennas are strongly correlated. Define the 
covariance of vectorized channel ${\bf h} = {\rm vec}(\mb{H}$) as 
${\mathsf 
E}\left( 
{{\bf{h}}{{\bf{h}}^H}} 
\right) = \bm{\Sigma}_{\bf h}\in{\mathbb C}^{N_{\rm R}N_{\rm T}\times N_{\rm 
R}N_{\rm T}}$,  which is also called the {\it kernel} of channel. The 
high-correlation property of 
wireless channel indicates that the kernel $\bm{\Sigma_{\bf h}}$ can provide prior knowledge to achieve 
accurate channel estimation \cite{werner2009estimating,Sungwoo'18,Karthik'18,Khalilsarai;20}. To this end, we follow the 
idea of GPR to design the estimator and 
the observation matrix. According to the model in (\ref{eq:channel_model}), the channel is 
sampled from the Gaussian process ${\cal 
	CN}\left({\bf 0}_{N_{\rm R}N_{\rm T}}, {\bm \Sigma}_{\bf h}\right)$.
The joint probability 
distribution of $\mb{h}$ and $\mb{y}$ then satisfies
\begin{equation}
	\begin{aligned}
		\left[ \begin{array}{l}
			{\bf{h}}\\
			{{\bf{y}}}
		\end{array} \right] \sim \mathcal{CN} \left(
		\left[ \begin{array}{c}
			{\bf 0}_{N_{\rm R}N_{\rm T}} \\
			{\bf 0}_{N_{\rm RF}Q}
		\end{array} \right], 
		\left[ 
		{\begin{array}{*{20}{c}}
				{{{\bf{\Sigma }}_{\bf{h}}}}&{{\bf{\Sigma 
					}}_{\bf{h}}}{{\bf{X}}}\\{{\bf{X}}^H}
				{{\bf{\Sigma 
					}}_{\bf{h}}}& {\bf X}^{H}{\bf{\Sigma 
				}}_{\bf{h}}{\bf X} + {\bm \Xi}
		\end{array}} \right]\right),
	\end{aligned}
\end{equation}
where ${\bm \Xi} = \sigma^2{\rm blkdiag}\left( 
{ {\bf W}_1^H{\bf W}_1, \cdots ,{\bf W}_Q^H{\bf W}_Q} \right)$ represents 
the covariance matrix of 
the noise $\bf z$. Then,  
the posterior mean and the posterior covariance 
of $\bf h$ are expressed as
\begin{equation}\label{eq:postmean}
	{{\bm{\mu }}}_{\mb{h}|\mb{y}} = {{\bf{\Sigma 
		}}_{\bf{h}}}{{\bf{X}}}
	{\left( {\bf X}^{H}{\bf{\Sigma 
		}}_{\bf{h}}{\bf X} + {\bm \Xi} \right)^{ - 
			1}} {{\bf{y}}},
\end{equation}
\begin{equation}
	{{\bf{\Sigma }}}_{\mb{h}|\mb{y}} = {{\bf{\Sigma 
		}}_{\bf{h}}} -
	{{\bf{\Sigma 
		}}_{\bf{h}}}{{\bf{X}}}
	{\left( {\bf X}^{H}{\bf{\Sigma 
		}}_{\bf{h}}{\bf X} + {\bm \Xi} \right)^{ - 
			1}}{ 
		{{{\bf{X}}^H}} }
	{{\bf{\Sigma }}_{\bf{h}}}.
\end{equation}
The posterior mean ${{\bm{\mu 
}}}_{\mb{h}|\mb{y}}$ is exactly the adopted channel estimator, which is equivalent to the linear MMSE (LMMSE) estimator. Besides, 
the posterior covariance ${{\bf{\Sigma }}}_{\mb{h}|\mb{y}}$ characterizes the estimation error, which is largely 
dependent on the observation matrix $\mb{X}$. This dependence indicates that well-designed 
combiners and precoders, $\{\mb{W}_q\}_{q=1}^Q$ and 
$\{\mb{v}_q\}_{q=1}^Q$, can 
substantially reduce the channel estimation error. Motivated by this fact, 
GPR attempts to produce the observation matrices to gain as much information 
of $\mb{h}$ as possible from the received signal $\mb{y}$.   
Henceforth, our 
objective is to find $\{\mb{W}_q\}_{q=1}^Q$ and $\{\mb{v}_q\}_{q=1}^Q$ that 
maximize the MI between $\mb{y}$ and $\mb{h}$, which is formulated as:
\begin{align}\label{eq:Entropy}
	\max_{\mb{W}\in{\cal W},\|\mb{v}_q\|^2 = P}~I(\mb{y}; \mb{h}) 
	= 
	\log\det\left( \mb{I}_{N_{\rm RF}Q} +  {\bf 
		\Xi}^{-1}\mb{X}^H{\bf 
		\Sigma_h} \mb{X} \right).
\end{align}
where $\mathcal{W}$ stands for the feasible set of hybrid combiners depending 
on the receiver hardware. In the subsequent
sections, we will first elaborate on the observation matrix design while 
considering the amplitude-and-phase controllable analog combiners 
 in Section \ref{sec:2DIF}. Then, our design will be extended to the case of 
 phase-only controllable analog combiners in Section \ref{sec:TS-2DIF}.

\section{Proposed Two-Dimensional Ice Filling (2DIF) Based Observation Matrix Design}\label{sec:2DIF}
In this section, we consider the ideal case when the amplitudes and phases of 
all precoder and combiner coefficients are adjustable, as shown in 
Fig. \ref{img:system} (a). In this 
context, $\{\mb{W}_q\}_{q=1}^Q$ can be freely configured and 
$\{\mb{v}_q\}_{q=1}^Q$ should satisfy the transmit power constraints $\left\| 
{{{\bf{v}}_q}} \right\|^2 = P$ for all $ q\in\{1,\cdots,Q\}$.

\subsection{Precoder/Combiner Design Using Greedy Method}\label{subsec:PCD_Greedy}
Observing problem (\ref{eq:Entropy}), one can find that the MI $I(\mb{y}; 
\mb{h})$ is non-concave with respect to (w.r.t) the overall observation matrix 
$\bf X$. Besides, due to the coupled term ${{\bf{v}}_q^T \otimes {\bf{W}}_q^H}$ 
in $\bf X$ and the colored-noise covariance matrix $\bf \Xi$ introduced by 
combiners $\{\mb{W}_q\}_{q=1}^Q$, the global optimal solution to 
(\ref{eq:Entropy}) is 
hard to obtain. To address this issue, we adopt a greedy method to generate 
$\{\mb{W}_q\}_{q=1}^Q$ and $\{\mb{v}_q\}_{q=1}^Q$ in a pilot-by-pilot manner. 
Specifically, we define ${{\bf{\bar X}}_t} = [{\bf{X}}_1, 
{\bf{X}}_2, \cdots, {\bf{X}}_t]$ as the overall observation 
matrix for timeslots 
$1\sim t$, where $t \le Q$. Let ${\bf{\bar y}}_{t} 
= {\bf{\bar X}}_{t}^H\mb{h} + {\bf{\bar z}}_{t}$ denote the corresponding 
received signal, wherein ${{\bf{\bar y}}_t} = [{\bf{y}}_1^T, 
{\bf{y}}_2^T, \cdots, {\bf{y}}_t^T]^T$ and ${\bf{\bar z}}_{t} := \left[ {\bf 
z}_1^H{\bf 
	W}_1,\cdots, {\bf z}_t^H{\bf W}_t \right]^H$. 
Given the current observation 
matrices $\{\mb{W}_q\}_{q=1}^t$ and vectors $\{\mb{v}_q\}_{q=1}^t$ in the first $t$ timeslots, our greedy strategy aims to find the 
combiner $\mb{W}_{t+1}$ and the precoder $\mb{v}_{t+1}$ in the next timeslot, which maximize the MI increment  
from timeslot $t$ to $t+1$: 
\begin{align}\label{eq:MI_incre_Max}
\mathop {\max }\limits_{\mb{W}_{t+1},\mb{v}_{t+1}}~
\Delta I_{t+1} := I({\bf{\bar y}}_{t+1};\mb{h}) - 
I({\bf{\bar y}}_t; \mb{h}). 
\end{align}
For clarity, we summarize the proposed design strategy in {\bf Algorithm 1}, and the sequential designs of $\{\mb{W}_q\}_{q=1}^Q$ and $\{\mb{v}_q\}_{q=1}^Q$ are illustrated as follows.

\begin{algorithm}[!t]
	\caption{2DIF Based Combiner and Precoder Design} 
	\begin{algorithmic}[1]\label{alg:2DIF}
		\REQUIRE  
		Number of pilots $Q$, kernel ${\bm \Sigma}_{\bf h}$.
		\ENSURE 
		Designed precoders $\{\mb{v}^{\rm opt}_q\}_{q=1}^Q$ and combiners $\{\mb{W}^{\rm opt}_q\}_{q=1}^Q$.
		\STATE Rewrite kernel as ${\bf{\Sigma }}_{\bf{h}} = {\bf{\Sigma }}_{\rm T} \otimes  {\bf{\Sigma }}_{\rm R}$
		\STATE Find the eigenvectors $[\mb{a}_1, \mb{a}_2, \cdots, \mb{a}_{N_{\rm T}}]$ 
		and the corresponding eigenvalues $[\alpha_1, \alpha_2, \cdots, 
		\alpha_{N_{\rm T}}]$ of ${\bf{\Sigma }}_{\rm T}$ 
		\STATE Find the eigenvectors $[\mb{b}_1, \mb{b}_2, \cdots, \mb{b}_{N_{\rm R}}]$ 
		and the corresponding eigenvalues $[\beta_1, \beta_2, \cdots, 
		\beta_{N_{\rm R}}]$ of ${\bf{\Sigma }}_{\rm R}$ 	
		\STATE Initialize: $[\lambda _{1,1}^0,\lambda _{1,2}^0, \cdots ,\lambda _{{N_{\rm{T}}},{N_{\rm{R}}}}^0] = [{\alpha _1}{\beta _1},{\alpha _1}{\beta _2}, \cdots ,{\alpha _{{N_{\rm{T}}}}}{\beta _{{N_{\rm{R}}}}}]$
		\FOR{$t = 0, \cdots, Q-1$}
		\STATE Find the optimal $n_{\rm T}^{\rm opt}$ and $\{ {{n^{\rm opt}_{{\rm{R}},k}}} \}_{k = 1}^{{N_{{\rm{RF}}}}}$ via {\bf Algorithm 2}
		\STATE Eigenvector-assignment: 	${\bf{v}}_{t + 1}^ {\rm opt}  = \sqrt{P} 
		{\bf{a}}_{n_{\rm{T}}^ {\rm opt} }^*$ and ${\bf{W}}_{t + 1}^ {\rm opt}  = 
		[{{{\bf{b}}_{n_{{\rm{R}},1}^ {\rm opt} }}, \cdots 
		,{{\bf{b}}_{n_{{\rm{R}},{N_{{\rm{RF}}}}}^ {\rm opt} }}}]$
		\STATE Eigenvalue-update for all ${n_{\rm{T}}} \in \left\{ {1, \cdots ,{N_{\rm{T}}}} \right\}$ and ${n_{\rm{R}}} \in \left\{ {1, \cdots ,{N_{\rm{R}}}} \right\}$ via
		\STATE ${\lambda^{t + 1}_{{n_{\rm{T}}},{n_{\rm{R}}}}} = \left\{ {\begin{matrix}
				\frac{{\lambda^t_{{n_{\rm{T}}},{n_{\rm{R}}}}}{\sigma ^2}} 
				{P\lambda^t_{{n_{\rm{T}}},{n_{\rm{R}}}} + \sigma ^2}, &  \!\!\! 
				n_{\rm{T}} \!=\! n^{\rm opt}_{\rm{T}}~\&~n_{{\rm{R}}}\in\{ 
				{{n^{\rm opt}_{{\rm{R}},k}}} \}_{k = 1}^{{N_{{\rm{RF}}}}}, \cr 
				{{\lambda^t_{{n_{\rm{T}}},{n_{\rm{R}}}}}}, & {{\text{else}}}.  \cr 
		\end{matrix} } \right.$
		\ENDFOR
		\RETURN Designed precoders $\{\mb{v}^{\rm opt}_q\}_{q=1}^Q$ and combiners $\{\mb{W}^{\rm opt}_q\}_{q=1}^Q$ for channel estimation.
	\end{algorithmic}
\end{algorithm}

\subsubsection{When $t=1$}
For ease of understanding, we begin with handling the first timeslot, i.e., 
$t=1$. In this context, problem (\ref{eq:MI_incre_Max}) can be rewritten as
\begin{align}
	\label{eq:problem_t_1}
\mathop {\max }\limits_{ \|{\bf{v}}_1\|^2 = P,\:{\bf W}_1}~
	&I(\mb{y}_{1};\mb{h}) 
\end{align}	
where the mutual information $I(\mb{y}_{1};\mb{h})$ is given by
\begin{align}
\notag
I(\mb{y}_{1};\mb{h}) =& {\log}\det {\bigg  (}  {{\bf{I}}_{{N_{{\rm{RF}}}}}} + 
		\frac{1}{\sigma^2 }{{\left( {{\bf{W}}_1^H{{\bf{W}}_1}} \right)}^{-1}}\times
		\\
		& \left( 
		{{\bf{v}}_1^T \otimes {\bf{W}}_1^H} \right){{\bf{\Sigma }}_{\bf{h}}}\left( 
		{{\bf{v}}_1^* \otimes {\bf{W}}_1} \right) {\bigg  )}.
		\label{eq:MI_1}
\end{align}
Since the reformulated problem (\ref{eq:problem_t_1}) is still 
intricate, we seek to simplify it using the following lemmas. 

\begin{lemma}\label{lemma:Sigma_decomp}
For the correlated Rayleigh-fading model in (\ref{eq:channel_model}), the covariance of the vectored channel ${\bf{\Sigma 
}}_{\bf{h}}$ can be rewritten as the form of a Kronecker product of two 
kernels, i.e.,
\begin{equation}
{\bf{\Sigma }}_{\bf{h}} = {\bf{\Sigma }}_{\rm T} \otimes  {\bf{\Sigma }}_{\rm R},
\end{equation}
where ${\bf{\Sigma }}_{\rm T}: = {{{( {{\bf{C}}_{{\rm{tx}}}^{1/2}} )}^T}{\bf{R}}_{{\rm{tx}}}^*{{( {{\bf{C}}_{{\rm{tx}}}^{1/2}} )}^*}} \in{\mathbb C}^{N_{\rm T}\times N_{\rm T} }$ and 
${\bf{\Sigma }}_{\rm R}:= {{\bf{C}}_{{\rm{rx}}}^{1/2}{\bf{R}}_{{\rm{rx}}}{{( {{\bf{C}}_{{\rm{rx}}}^{1/2}})}^H}} \in{\mathbb C}^{N_{\rm R}\times N_{\rm R}}$ characterize 
the channel correlation among transmit antennas and that among receive 
antennas, respectively.
\end{lemma}
\begin{IEEEproof}
See Appendix \ref{appendix:Sigma_decomp}.
\end{IEEEproof}
\begin{lemma}\label{lemma:Orth_W}
Introducing the orthogonality constraints ${\bf{W}}_q^H{{\bf{W}}_q} = 
{{\bf{I}}_{N_{\rm RF}}}$ for all $q\in\{1,\cdots,Q\}$ does not influence the 
optimal value of MI $I({\bf{y}}; \mb{h})$ in (\ref{eq:Entropy}).
\end{lemma}
\begin{IEEEproof}
See Appendix \ref{appendix:Orth_W}.
\end{IEEEproof}

Utilizing {\bf Lemma \ref{lemma:Sigma_decomp}} and {\bf Lemma 
\ref{lemma:Orth_W}}, problem (\ref{eq:problem_t_1}) can be equivalently 
rewritten as 
\begin{align}\label{eq:problem_t_1_refor}
	\notag
	\mathop {\max }\limits_{ {\bf{v}}_1, {\bf W}_1}~
	&I(\mb{y}_{1};\mb{h})= {\log}\det \left( {{{\bf{I}}_{{N_{{\rm{RF}}}}}} \!+\! {{{\bf{v}}_1^T{{\bf{\Sigma }}_{\rm{T}}}{\bf{v}}_1^*} \over {{\sigma ^2}}} {{\bf{W}}_1^H{{\bf{\Sigma }}_{\rm{R}}}{{\bf{W}}_1}}} \right) \\
	\notag
	{\rm s.t.}~~&\left\| {{{\bf{v}}_1}}\right\|^2 = P, 
	\\
	&{\bf W}_1^H{\bf W}_1 = {\bf I}_{N_{\rm RF}},
\end{align}	
where the reorganized objective function $I(\mb{y}_{1};\mb{h})$ is obtained by 
substituting ${\bf{\Sigma }}_{\bf{h}} = {\bf{\Sigma }}_{\rm T} \otimes  
{\bf{\Sigma }}_{\rm R}$ and ${\bf W}_1^H{\bf W}_1 = {\bf I}_{N_{\rm RF}}$ into 
(\ref{eq:problem_t_1}). 
Observing (\ref{eq:problem_t_1_refor}), 
one can prove with ease that the 
optimal ${\bf v}_1^*$ is the eigenvector of ${\bf{\Sigma }}_{\rm{T}}$ 
associated with its largest eigenvalue, and the optimal ${\bf W}_{1}$ is composed of the 
eigenvectors of ${\bf{\Sigma }}_{\rm{R}}$ associated with its top $N_{\rm 
RF}$ eigenvalues. Define the eigenvalue decompositions (EVDs) ${\bf{\Sigma 
}}_{\rm{T}} = {{\bf{U}}_{\rm{T}}}{{\bf{\Lambda 
}}_{\rm{T}}}{{\bf{U}}_{\rm{T}}^H}$ and ${{\bf{\Sigma }}_{\rm{R}}} = 
{{\bf{U}}_{\rm{R}}}{{\bf{\Lambda}}_{\rm{R}}}{\bf{U}}_{\rm{R}}^H$, wherein the 
eigenvalues in ${{\bf{\Lambda }}_{\rm{T}}} = {\rm{diag}}\left( {{\alpha _1}, 
\cdots ,{\alpha _{{N_{\rm{T}}}}}} \right)$ and ${{\bf{\Lambda }}_{\rm{R}}} = 
{\rm{diag}}\left( {{\beta _1}, \cdots ,{\beta _{{N_{\rm{R}}}}}} \right)$ are 
arranged in descending order. In this way, the maximum MI in 
(\ref{eq:problem_t_1_refor}) can be derived as
\begin{align}\label{eq:optimal_solution_1}
\mathop {\max }\limits_{ {\bf{v}}_1, {\bf W}_1}~
I(\mb{y}_{1};\mb{h}) = \sum\limits_{n = 1}^{{N_{{\rm{RF}}}}} {\log_2\left( {1 + 
{{P{\alpha _1}{\beta _n}} \over {{\sigma ^2}}}} \right)},
\end{align}
and its achievable solution are expressed as
\begin{align}\label{eq:optimal_v_W_1}
{{\bf{v}}_1^{\rm opt}} = \sqrt{P}{\bf{U}}_{\rm{T}}^*\left( :,1 \right)~\text{and}~
{{\bf{W}}_1^{\rm opt}} = {{\bf{U}}_{\rm{R}}}\left( {:,\left[ {1,\cdots,{N_{{\rm{RF}}}}} 
\right]} \right).
\end{align}
Equation \eqref{eq:optimal_v_W_1} can be viewed as the optimal initialization 
settings for our proposed observation matrix design. 
 
\subsubsection{From $t$ to $t+1$}
Given the optimized precoder ${\bf v}_1$ and combiner ${\bf W}_1$ at timeslot 
$t=1$, 
we then consider designing $\{\mb{v}_q\}_{q=2}^Q$ and $\{\mb{W}_q\}_{q=2}^Q$ in 
a sequential manner. By invoking the principle of recursion, we only need to 
address 
the design of precoder ${\bf v}_{t+1}$ and combiner ${\bf W}_{t+1}$ with given 
$\{\mb{W}_q\}_{q=1}^{t}$ and $\{\mb{v}_q\}_{q=1}^{t}$. As proved in Appendix 
\ref{appendix:MI_incre}, the MI increment, $\Delta I_{t+1}$, can be 
equivalently rewritten as
\begin{align}
\Delta I_{t+1} = {\log _2}\det \left( {{{\bf{I}}_{{N_{{\rm{RF}}}}}} \!+\! {1 
\over 
 {{\sigma ^2}}}{\bf{X}}_{t + 1}^H{{\bf{\Sigma }}_{t}}{{\bf{X}}_{t + 1}}} 
 \right),
\end{align}
wherein ${\bf X}_{t+1}:={{\bf{v}}_{t+1}^* \otimes {\bf{W}}_{t+1}}$ refers to the 
Kronecker-constrained observation matrix, and 
\begin{align}\label{eq:Sigma_t}
{{\bf{\Sigma }}_t} = {{\bf{\Sigma }}_{\bf{h}}} - {{\bf{\Sigma }}_{\bf{h}}}{{{\bf{\bar X}}}_t}{\left( {{\bf{\bar X}}_t^H{{\bf{\Sigma }}_{\bf{h}}}{{{\bf{\bar X}}}_t} + {\sigma ^2}{{\bf{I}}_{{N_{{\rm{RF}}}}t}}} \right)^{ - 1}}{\bf{\bar X}}_t^H{{\bf{\Sigma }}_{\bf{h}}}
\end{align}
is the posterior kernel of channel $\bf h$ given the observation ${{\bf{\bar y}}}_t$. In particular, we have ${{\bf{\Sigma }}_0}:={{\bf{\Sigma }}_{\bf h}}$. Utilizing {\bf Lemma \ref{lemma:Sigma_decomp}} and {\bf Lemma \ref{lemma:Orth_W}}, the optimal precoder ${\bf v}_{t+1}$ and combiner ${\bf W}_{t+1}$ at the $(t+1)$-th timeslot can be obtained by solving
\begin{align}\label{eq:problem_t+1}
	\notag
	\mathop {\max }\limits_{ {\bf{v}}_{t+1}, {\bf W}_{t+1}}
	&\log\det \left( {{\bf{I}}_{{N_{{\rm{RF}}}}}} \!+\! {1 \over {{\sigma ^2}}}({\bf v}_{t+1}^T\otimes \mb{W}_{t+1}^H){{\bf{\Sigma }}_{t}}({\bf v}_{t+1}^*\otimes \mb{W}_{t+1}) \right) \\
	\notag
	{\rm s.t.}~~&\left\| {{{\bf{v}}_{t+1}}}\right\|^2 = P, 
	\\
	&{\bf W}_{t+1}^H{\bf W}_{t+1} = {\bf I}_{N_{\rm RF}}.
\end{align}

Note that, in problem (\ref{eq:problem_t_1_refor}), the kernel ${\bf{\Sigma 
}}_{\bf{h}}$ is decoupled into ${\bf{\Sigma }}_{\rm T} \otimes  {\bf{\Sigma 
}}_{\rm R}$ such that ${{\bf{v}}_1}$ and ${{\bf{W}}_1}$ can be obtained by 
selecting the appropriate eigenvectors of ${\bf{\Sigma }}_{\rm T}$ and 
${\bf{\Sigma }}_{\rm R}$ as in (\ref{eq:optimal_v_W_1}). We 
attempt the similar idea to solve for ${{\bf{v}}_{t+1}}$ and ${{\bf{W}}_{t+1}}$. 
To this end, we first define the EVD: ${\bf{\Sigma }}_{t} = 
{{\bf{U}}_{t}}{{\bf{\Lambda }}_{t}}{{\bf{U}}_{t}^H}$ where 
${\bf{U}}_{t}\in{\mathbb 
C}^{{N_{\rm{T}}}{N_{\rm{R}}}\times{N_{\rm{T}}}{N_{\rm{R}}}}$.
Notice that the 
constraints $\left\| {{{\bf{v}}_{t+1}}}\right\|^2 = P$ and ${\bf W}_{t+1}^H{\bf 
W}_{t+1} = {\bf I}_{N_{\rm RF}}$ make the overall matrix observation ${\bf 
X}_{t+1}$ 
orthogonal, i.e., ${\bf X}_{t+1}^H{\bf X}_{t+1} = P\mb{I}_{\rm RF}$. 
If we 
temporarily omit the Kronecker constraint ${\bf X}_{t+1}={{\bf{v}}_{t+1}^* 
\otimes {\bf{W}}_{t+1}}$ and try to solve \eqref{eq:problem_t+1} by considering 
the orthogonal constraint ${\bf X}_{t+1}^H{\bf X}_{t+1} = P\mb{I}_{\rm RF}$ 
only, it becomes evident that the global optimal solution to 
\eqref{eq:problem_t+1} 
is ${\bf 
X}_{t+1}= \sqrt{P}{{\bf{U}}_{t}}\left( {:,\left[{1,\cdots,{N_{{\rm{RF}}}}} 
\right]} \right)$. Motivated by this discovery, a natural question arises: 
\emph{when the Kronecker constraint holds, is it possible to set ${\bf 
X}_{t+1}$ as the 
principal eigenvectors of ${\bf{\Sigma }}_{t}$ by properly designing 
${{\bf{v}}_{t+1}}$ and ${\bf 
W}_{t+1}$?}
Addressing this question is crucial for generating near-optimal observation 
matrices. We would like to investigate it by analyzing the impacts and 
feasibility 
of setting ${\bf X}_{t+1}$ as the principal eigenvectors of ${\bf{\Sigma 
}}_{t}$.

 \emph{ i) Influence of setting  ${\bf 
		X}_{t+1}$ as the 
	principal eigenvectors of ${\bf{\Sigma }}_{t}$.} Before evaluating the 
	feasibility of setting ${\bf 
		X}_{t+1}= \sqrt{P}{{\bf{U}}_{t}}\left( 
		{:,\left[{1,\cdots,{N_{{\rm{RF}}}}} 
		\right]} \right)$, 
we first need to exploit its influence on  the 
	evolution rule of the posterior kernel  ${\bf{\Sigma }}_{t + 1}$.
The following 
lemma characterizes the relationship between ${\bf{\Sigma }}_{t + 1}$ 
and ${\bf{\Sigma }}_{t}$. 
\begin{lemma}\label{lemma:Sigma_t_lambda}
Let ${\lambda_n}\left(\cdot\right)$ denote the $n$-th largest eigenvalue of the 
matrix in its argument, e.g., ${\lambda 
	_n}\left( {{{\bf{\Sigma }}_{t}}} \right) = {{\bf{\Lambda }}_{t}}\left( 
{n,n} \right)$ for ${\bf{\Sigma }}_{t} = 
{{\bf{U}}_{t}}{{\bf{\Lambda }}_{t}}{{\bf{U}}_{t}^H}$. If 
${{\bf{X}}_{t+1}} = \sqrt{P}{{\bf{U}}_{t}}\left( {:,[1, \cdots 
	,{N_{{\rm{RF}}}}]} 
\right)$,
then the EVD of ${{\bf{\Sigma }}_{t+1}}$ can be derived 
from ${\bf{\Sigma }}_{t} = 
{{\bf{U}}_{t}}{{\bf{\Lambda }}_{t}}{{\bf{U}}_{t}^H}$ by
\begin{align}\label{eq:Sigma_t_U_t}
	\notag
	{{\bf{\Sigma }}_{t+1}} =& {{\bf{U}}_{t}}{\rm{diag}}{\bigg (} 
	{{\lambda_1\left( 
			{{{\bf{\Sigma }}_{t}}} \right)\sigma^2} \over {P{\lambda 
			_1}\left( 
			{{{\bf{\Sigma }}_{t}}} \right) + {\sigma ^2}}}, {{{\lambda 
			_2\left( 
				{{{\bf{\Sigma }}_{t}}} \right)\sigma^2} \over {P{\lambda 
				_2}\left( 
				{{{\bf{\Sigma }}_{t}}} \right) + {\sigma ^2}}}}, \cdots, \\ 
				\notag & 
	{{\lambda _{{N_{{\rm{RF}}}}}\left( {{{\bf{\Sigma }}_{t}}} 
	\right)\sigma^2} 
		\over {P{\lambda _{{N_{{\rm{RF}}}}}}\left( {{{\bf{\Sigma }}_{t}}} 
		\right) + 
			{\sigma ^2}}},{\lambda _{{N_{{\rm{RF}}}} + 1}}\left( {{{\bf{\Sigma 
			}}_{t}}} 
	\right),{\lambda _{{N_{{\rm{RF}}}} + 2}}\left( {{{\bf{\Sigma }}_{t}}} 
	\right), \\ &\cdots ,{\lambda _{{N_{\rm{R}}}{N_{\rm{T}}}}}\left( 
	{{{\bf{\Sigma 
			}}_{t}}} \right) {\bigg )}{{\bf{U}}_{t}^H}.
\end{align}
\end{lemma}
\begin{IEEEproof}
See Appendix \ref{appendix:Sigma_t_lambda}.
\end{IEEEproof}

{\bf Lemma \ref{lemma:Sigma_t_lambda}} reveals that, when ${{\bf{X}}_{t+1}} = 
\sqrt{P}{{\bf{U}}_{t}}\left( {:,[1, \cdots ,{N_{{\rm{RF}}}}]} \right)$, the 
posterior covariance matrix ${{\bf{\Sigma }}_{t+1}}$ shares the same 
eigenvector space, ${{\bf{U}}_{t}}$, 
as ${{\bf{\Sigma }}_{t}}$. The only difference on their EVDs is that the 
$N_{\rm RF}$-largest 
eigenvalues of ${{\bf{\Sigma }}_{t}}$, i.e., $\left\{ {{\lambda 
	_n}({{\bf{\Sigma }}_{t}})} \right\}_{n = 1}^{{N_{{\rm{RF}}}}}$, are 
replaced by $\{ {{{\lambda _n({{\bf{\Sigma }}_{t}})\sigma^2} \over 
	{P{\lambda _n}({{\bf{\Sigma }}_{t}}) + {\sigma ^2}}}} \}_{n = 
1}^{{N_{{\rm{RF}}}}}$ in ${{\bf{\Sigma }}_{t + 1}}$. Considering the generality 
of 
$t$, we can conclude that, 
the eigenvectors of channel covariance ${{\bf{\Sigma }}_0}:={{\bf{\Sigma 
}}_{\bf h}}$ are preserved by all subsequent posterior kernels ${{\bf{\Sigma 
}}_1}$, ${{\bf{\Sigma }}_2}$, $\cdots$, and ${{\bf{\Sigma }}_{Q-1}}$. In other 
words, the only difference among ${{\bf{U}}_0}$, ${{\bf{U}}_1}$, $\cdots$, 
${{\bf{U}}_{Q - 1}}$ is their column arrangement orders. 
	
 \emph{ ii) Feasibility of setting  ${\bf 
		X}_{t+1}$ as the 
	principal eigenvectors of ${\bf{\Sigma }}_{t}$.}
Given the eigenvector-preserving property in {\bf Lemma \ref{lemma:Sigma_t_lambda}}, the feasibility of setting 
${\bf X}_{t+1}$ as the principal eigenvectors of ${\bf{\Sigma }}_{t}$ lies in 
the feasibility of setting ${\bf X}_{t+1}$ as the eigenvectors of ${\bf{\Sigma 
}}_{0} = {\bf{\Sigma 
}}_{\mb{h}}$. To assess this feasibility, we examine the structure of the  
eigenspace of ${{\bf{\Sigma }}_{\bf h}}$ below. 
\begin{corollary}\label{corollary:Sigma_h_aT_bR}
The EVD of the prior covariance ${{\bf{\Sigma }}_{\bf h}}$ can be written as
\begin{align}\label{eq:Sigma_h_T_R}
\notag
{{\bf{\Sigma }}_{\bf{h}}} = & {{\bf{U}}_0}{{\bf{\Lambda }}_0}{\bf{U}}_0^H
\\ \notag
= & \underbrace {\left( {{{\bf{U}}_{\rm{T}}} \otimes {{\bf{U}}_{\rm{R}}}} 
\right)}_{\text{Orthogonal matrix}}\underbrace {\left( {{{\bf{\Lambda 
}}_{\rm{T}}} \otimes {{\bf{\Lambda }}_{\rm{R}}}} \right)}_{\text{Eigenvalue 
matrix}}\left( {{\bf{U}}_{\rm{T}}^H \otimes {\bf{U}}_{\rm{R}}^H} \right) \\
= &  \sum\limits_{{n_{\rm{T}}} = 1}^{{N_{\rm{T}}}} {\sum\limits_{{n_{\rm{R}}} = 1}^{{N_{\rm{R}}}} {{\alpha _{{n_{\rm{T}}}}}{\beta _{{n_{\rm{R}}}}}\left( {{{\bf{a}}_{{n_{\rm{T}}}}} \otimes {{\bf{b}}_{{n_{\rm{R}}}}}} \right){{\left( {{{\bf{a}}^H_{{n_{\rm{T}}}}} \otimes {{\bf{b}}^H_{{n_{\rm{R}}}}}} \right)}}} },
\end{align}
where ${{\bf{a}}_{{n_{\rm{T}}}}} := {{\bf{U}}_{\rm{T}}}\left( {:,{n_{\rm{T}}}} \right)$ denotes the $n_{\rm{T}}$-th eigenvector of ${\bf{\Sigma }}_{\rm{T}}$ and ${{\bf{b}}_{{n_{\rm{R}}}}} := {{\bf{U}}_{\rm{R}}}\left( {:,{n_{\rm{R}}}} \right)$ denotes the $n_{\rm{R}}$-th eigenvector of ${\bf{\Sigma }}_{\rm{R}}$. In particular, $\left\{ {{\alpha _{{n_{\rm{T}}}}}{\beta _{{n_{\rm{R}}}}}} \right\}_{{n_{\rm{T}}} = 1,{n_{\rm{R}}} = 1}^{{N_{\rm{T}}},{N_{\rm{R}}}}$ and $\left\{ {{{\bf{a}}_{{n_{\rm{T}}}}} \otimes {{\bf{b}}_{{n_{\rm{R}}}}}} \right\}_{{n_{\rm{T}}} = 1,{n_{\rm{R}}} = 1}^{{N_{\rm{T}}},{N_{\rm{R}}}}$ are the eigenvalues and the corresponding eigenvectors of ${{\bf{\Sigma }}_{\bf{h}}}$, respectively.
\end{corollary}
\begin{IEEEproof}
See Appendix \ref{appendix:Sigma_h_aT_bR}.
\end{IEEEproof}

Recalling that ${\bf X}_{t + 1}:={{\bf{v}}_{t + 1}^* \otimes {\bf{W}}_{t + 
1}}$, we find that the analytical form of ${\bf X}_{t + 1}$ perfectly matches 
that 
of the eigenvectors of ${{\bf{\Sigma }}_{\bf{h}}}$, i.e., $\left\{ 
{{{\bf{a}}_{{n_{\rm{T}}}}} \otimes {{\bf{b}}_{{n_{\rm{R}}}}}} 
\right\}_{{n_{\rm{T}}} = 1,{n_{\rm{R}}} = 1}^{{N_{\rm{T}}},{N_{\rm{R}}}}$. This 
encouraging fact inspires us that, the desired optimal ${\bf X}_{t+1}$ to solve 
(\ref{eq:problem_t+1}) can be achieved by setting ${{\bf{v}}_{t+1}}$ and ${\bf 
W}_{t+1}$ to the  appropriate eigenvectors from $\left\{ 
\sqrt{P}{\bf{a}}^*_{{n_{\rm{T}}}} \right\}_{{n_{\rm{T}}} = 1}^{{N_{\rm{T}}}}$ 
and $\left\{{\bf{b}}_{{n_{\rm{R}}}} \right\}_{{n_{\rm{R}}} = 
1}^{{N_{\rm{R}}}}$, respectively. 

We provide an example to show the implementation process. Firstly, 
to achieve ${\bf X}_{1}= \sqrt{P}{{\bf{U}}_{0}}\left( {:,\left[ 
{1,\cdots,{N_{{\rm{RF}}}}} \right]} \right)$ at timeslot $t=1$, we can set 
${{\bf{v}}_1} = \sqrt{P}{\bf{a}}^*_1 = \sqrt{P}{\bf{U}}_{\rm{T}}^*\left( :,1 
\right)$ and 
${{\bf{W}}_1} = [{\bf{b}}_1,{\bf{b}}_2,\cdots,{\bf{b}}_{N_{\rm RF}}] = 
{{\bf{U}}_{\rm{R}}}\left( {:,\left[ {1,\cdots,{N_{{\rm{RF}}}}} \right]} 
\right)$, which coincides with the optimal solution in 
(\ref{eq:optimal_v_W_1}). Note that, according to {\bf Lemma 
\ref{lemma:Sigma_t_lambda}}, this process does not changes the eigenvectors 
of ${\bf \Sigma}_1$.\footnote{Thus, ${\bf{U}}_{1}$ and 
${\bf{U}}_{0}:={{{\bf{U}}_{\rm{T}}} \otimes {{\bf{U}}_{\rm{R}}}}$ share the 
same columns, but their column arrangement orders may be different due to the eigenvalue
updates.} In this context, the desired ${\bf X}_{2}= 
\sqrt{P}{{\bf{U}}_{1}}\left( {:,\left[ {1,\cdots,{N_{{\rm{RF}}}}} \right]} 
\right)$ at 
timeslot $t=2$ can also be achieved by carefully selecting ${{\bf{v}}_{2}}$ and 
${\bf W}_{2}$ from $\left\{ \sqrt{P}{\bf{a}}^*_{{n_{\rm{T}}}} 
\right\}_{{n_{\rm{T}}} = 
1}^{{N_{\rm{T}}}}$ and $\left\{{\bf{b}}_{{n_{\rm{R}}}} \right\}_{{n_{\rm{R}}} = 
1}^{{N_{\rm{R}}}}$ respectively, which does not influence the eigenvectors of 
${\bf \Sigma}_2$. Analogously, all our desired $\{{\bf X}_q\}^Q_{q=1}$ can be 
obtained by this successive process. As a result, the problem is transformed 
into: {\it How to select appropriate eigenvectors such that the objective 
function in problem (\ref{eq:problem_t+1}) is maximized?}

\subsection{Eigenvector Selection}\label{subsec:eig_selection}
In this subsection, the problem of eigenvector selection is investigated. According to {\bf Lemma \ref{lemma:Sigma_t_lambda}} and {\bf Corollary \ref{corollary:Sigma_h_aT_bR}}, all posterior kernels $\left\{ {{{\bf{\Sigma }}_t}} \right\}_{t = 0}^{Q - 1}$ can be rewritten as the form of 
\begin{align}\label{eq:Sigma_t_definition}
\notag
{{\bf{\Sigma }}_t} = &
{{\bf{U}}_0}{\rm{diag}}\left( {{\lambda^t_{1,1}},{\lambda^t_{1,2}}, \cdots ,{\lambda^t_{{N_{\rm{T}}},{N_{\rm{R}}}}}} \right){\bf{U}}_0^H
\\
= & \sum\limits_{{n_{\rm{T}}} = 1}^{{N_{\rm{T}}}} {\sum\limits_{{n_{\rm{R}}} = 1}^{{N_{\rm{R}}}} {{\lambda ^t_{{n_{\rm{T}}},{n_{\rm{R}}}}}\left( {{{\bf{a}}_{{n_{\rm{T}}}}} \otimes {{\bf{b}}_{{n_{\rm{R}}}}}} \right)\left( {{\bf{a}}_{{n_{\rm{T}}}}^H \otimes {\bf{b}}_{{n_{\rm{R}}}}^H} \right)} },
\end{align}
where ${{\lambda^t_{{n_{\rm{T}}},{n_{\rm{R}}}}}}$ is the $({{n_{\rm{T}}},{n_{\rm{R}}}})$-th eigenvalue of ${{\bf{\Sigma }}_t}$ associated with the eigenvector ${{{\bf{a}}_{{n_{\rm{T}}}}} \otimes {{\bf{b}}_{{n_{\rm{R}}}}}}$. In particular, ${\lambda^0_{{n_{\rm{T}}},{n_{\rm{R}}}}} = {\alpha _{{n_{\rm{T}}}}}{\beta _{{n_{\rm{R}}}}}$, and its update from $t$ to $t+1$ is expressed by
\begin{align}\label{eq:lambda_update}
{\lambda^{t + 1}_{{n_{\rm{T}}},{n_{\rm{R}}}}} = \left\{ {\begin{matrix}
		\dfrac{{\lambda^t_{{n_{\rm{T}}},{n_{\rm{R}}}}}{\sigma 
		^2}}{P\lambda^t_{{n_{\rm{T}}},{n_{\rm{R}}}} + \sigma ^2}, & 
		{{{\bf{a}}_{{n_{\rm{T}}}}^*} \otimes 
		{{\bf{b}}_{{n_{\rm{R}}}}}{\text{~is~selected,}}}  \cr 
		{{\lambda^t_{{n_{\rm{T}}},{n_{\rm{R}}}}}}, & {{\text{else}}}.  \cr 
\end{matrix} } \right.
\end{align}
Based on the above derivation, we prove the following lemma to further simplify the original problem (\ref{eq:problem_t+1}):
\begin{lemma}\label{lemma:eigen_selection}
When ${{\bf{v}}_{t+1}}\in\left\{ \sqrt{P}{\bf{a}}^*_{{n_{\rm{T}}}} 
\right\}_{{n_{\rm{T}}} = 1}^{{N_{\rm{T}}}}$ and 
the columns of $\mb{W}_{t+1}$ are belonging to 
$\left\{ {\bf{b}}_{{n_{\rm{R}}}} \right\}_{{n_{\rm{R}}} = 1}^{{N_{\rm{R}}}}$, 
the original problem 
(\ref{eq:problem_t+1}) can be transformed into an eigenvector 
selection problem, written as
\begin{align}\label{eq:problem_t+1_reorg}
	\notag
	\mathop {\max }\limits_{ {n_{\rm{T}}}, \left\{ {{n_{{\rm{R}},k}}} \right\}_{k = 1}^{{N_{{\rm{RF}}}}} }~
	&\sum\limits_{k = 1}^{{N_{{\rm{RF}}}}} {{{\log }_2}\left( {1 + 
	{{{P\lambda^t_{{n_{\rm{T}}},{n_{{\rm{R}},k}}}}} \over {{\sigma ^2}}}} 
	\right)}  \\
	\notag
	{\rm s.t.}~~~~~~&{n_{\rm{T}}} \in \left\{ {1, \cdots ,{N_{\rm{T}}}} \right\}, 
	\\
	\notag
	& {n_{{\rm{R}},{k}}} \in \left\{ {1, \cdots ,{N_{\rm{R}}}} \right\}, 
	\forall k,\\
	& {n_{{\rm{R}},k}} \ne {n_{{\rm{R}},k'}}, \forall k \ne k'.
\end{align}
\end{lemma}
\begin{IEEEproof}
See Appendix \ref{appendix:eigen_selection}.	
\end{IEEEproof}

\begin{algorithm}[!t]
	\caption{Linear Search for Eigenvalue Selection} 
	\begin{algorithmic}[1]\label{alg:TS_eig}
		\REQUIRE  
		Eigenvalues $\{ {{\lambda^t_{{n_{\rm{T}}},{n_{\rm{R}}}}}} \}_{{n_{\rm{T}}} = 1,{n_{\rm{R}}} = 1}^{{N_{\rm{T}}},{N_{\rm{R}}}}$ in timeslot $t$.
		\ENSURE 
		Optimal eigenvalue indices $n_{\rm T}^{\rm opt}$ and $\{ {{n^{\rm opt}_{{\rm{R}},k}}} \}_{k = 1}^{{N_{{\rm{RF}}}}}$ that maximize $\sum\nolimits_{k = 1}^{{N_{{\rm{RF}}}}} {{{\log }_2}( {1 + {{{\lambda^t_{{n_{\rm{T}}},{n_{{\rm{R}},k}}}}} / {{\sigma ^2}}}} )}$.
		\STATE Initialize indexes: $n_{\rm T}^{\rm opt}=1$ and $[n_{{\rm{R}},1}^ {\rm opt} , \cdots ,n_{{\rm{R}},{N_{{\rm{RF}}}}}^ {\rm opt} ] = [1, \cdots ,{N_{{\rm{RF}}}}]$.
		\STATE Initialize the maximum objective: $\zeta_{\max} = 
		\sum\nolimits_{k = 1}^{{N_{{\rm{RF}}}}} {{{\log }_2}( {1 + 
		{{{P\lambda^t_{{n^{\rm opt}_{\rm{T}}},{n^{\rm opt}_{{\rm{R}},k}}}}} / {{\sigma 
		^2}}}} )}$
		\FOR{$n_{\rm{T}}=1,\cdots,N_{\rm T}$}
		\STATE Find the $N_{\rm RF}$-largest values from $\{ {{\lambda^t_{{n_{\rm{T}}},{n_{\rm{R}}}}}} \}_{{n_{\rm{R}}} = 1}^{{N_{\rm{R}}}}$, and then denote their second indexes as $\{ {n_{{\rm{R}},k}}\} _{k = 1}^{{N_{{\rm{RF}}}}}$. 
		\IF{$\sum\nolimits_{k = 1}^{{N_{{\rm{RF}}}}} {{{\log }_2}( {1 + 
		{{{P\lambda^t_{{n_{\rm{T}}},{n_{{\rm{R}},k}}}}} / {{\sigma ^2}}}} 
		)}>\zeta_{\max}$}
		\STATE Update the optimal indexes by $n_{\rm T}^{\rm opt}=n_{\rm T}$ and $[n_{{\rm{R}},1}^ {\rm opt} , \cdots ,n_{{\rm{R}},{N_{{\rm{RF}}}}}^ {\rm opt} ] = [n_{{\rm{R}},1}, \cdots ,n_{{\rm{R}},{N_{{\rm{RF}}}}}]$
		\STATE Update the maximum objective: $\zeta_{\max} = 
		 \sum\nolimits_{k = 1}^{{N_{{\rm{RF}}}}} {{{\log }_2}( {1 + 
				{{{P\lambda^t_{{n^{\rm opt}_{\rm{T}}},{n^{\rm opt}_{{\rm{R}},k}}}}} / {{\sigma 
							^2}}}} )}$
		\ENDIF
		\ENDFOR
		\RETURN Optimal $n_{\rm T}^{\rm opt}$ and $\{ {{n^{\rm opt}_{{\rm{R}},k}}} \}_{k = 1}^{{N_{{\rm{RF}}}}}$.
	\end{algorithmic}
\end{algorithm}
Problem (\ref{eq:problem_t+1_reorg}) aims to find one eigenvector 
index $n_T$ on the transmitter side and $N_{\rm RF}$ different 
eigenvector indices $\left\{ 
{{n_{{\rm{R}},k}}} \right\}_{k = 1}^{{N_{{\rm{RF}}}}}$ on the receiver side, 
such that the selected eigenvalues $\{ 
{{\lambda^t_{{n_{\rm{T}}},{n_{{\rm{R}}, k}}}}} \}_{k = 1} 
	^{{N_{\rm{RF}}}}$ can maximize the objective 
	\eqref{eq:problem_t+1_reorg}. A linear search algorithm is proposed to 
	solve 
	(\ref{eq:problem_t+1_reorg}) optimally, as summarized in {\bf Algorithm 
	2}. 
	The 
key idea is to fix an ${n_{\rm{T}}}$ and then find $N_{\rm RF}$-largest values 
from $\{ {{\lambda^t_{{n_{\rm{T}}},{n_{\rm{R}}}}}} \}_{{n_{\rm{R}}} = 
1}^{{N_{\rm{R}}}}$ to calculate the objective $\sum\nolimits_{k = 
1}^{{N_{{\rm{RF}}}}} {{{\log }_2}( {1 + 
{P{{\lambda^t_{{n_{\rm{T}}},{n_{{\rm{R}},k}}}}} / {{\sigma ^2}}}} )}$. After 
traversing all ${n_{\rm{T}}}\in\{1,\cdots,N_{\rm T}\}$, the optimal $n_{\rm T}^{\rm opt}$ and $\{ 
{{n^{\rm opt}_{{\rm{R}},k}}} \}_{k = 1}^{{N_{{\rm{RF}}}}}$ can be obtained from the 
indices of the maximum objective. Finally, the desired precoder and 
combiner are thereby expressed as
\begin{align}\label{eq:optimal_v_W_t+1}
	{\bf{v}}_{t + 1}^ {\rm opt}  = \sqrt{P}{\bf{a}}_{n_{\rm{T}}^ {\rm opt} }^* 
	~\text{and}~ {\bf{W}}_{t + 1}^ {\rm opt}  = \left[{{{\bf{b}}_{n_{{\rm{R}},1}^ 
	{\rm opt} }}, \cdots ,{{\bf{b}}_{n_{{\rm{R}},{N_{{\rm{RF}}}}}^ {\rm opt} }}}\right],
\end{align}
respectively, which generate a feasible observation 
matrix 
$\mb{X}_{t+1}^{\rm opt} = {{\bf{v}}_{t + 1}^ {\rm opt}}^* \otimes {\bf{W}}_{t + 
1}^ {\rm opt}$ at timeslot $t + 1$. 
One can verify without difficulty that the initialized precoder and combiner 
obtained in \eqref{eq:optimal_v_W_1} are a special case of  
\eqref{eq:optimal_v_W_t+1} when $t = 0$. 

To summarize, the eigenvalue updating rule in \eqref{eq:lambda_update}, as well 
as the eigenvector selection method stated in \textbf{Algorithm 
2} 
and \eqref{eq:optimal_v_W_t+1}, allow us to calculate all observation 
matrices.

\subsection{Insightful Interpretation to 2DIF}

\begin{figure*}[!t]
	\centering
	\includegraphics[width=1\textwidth]{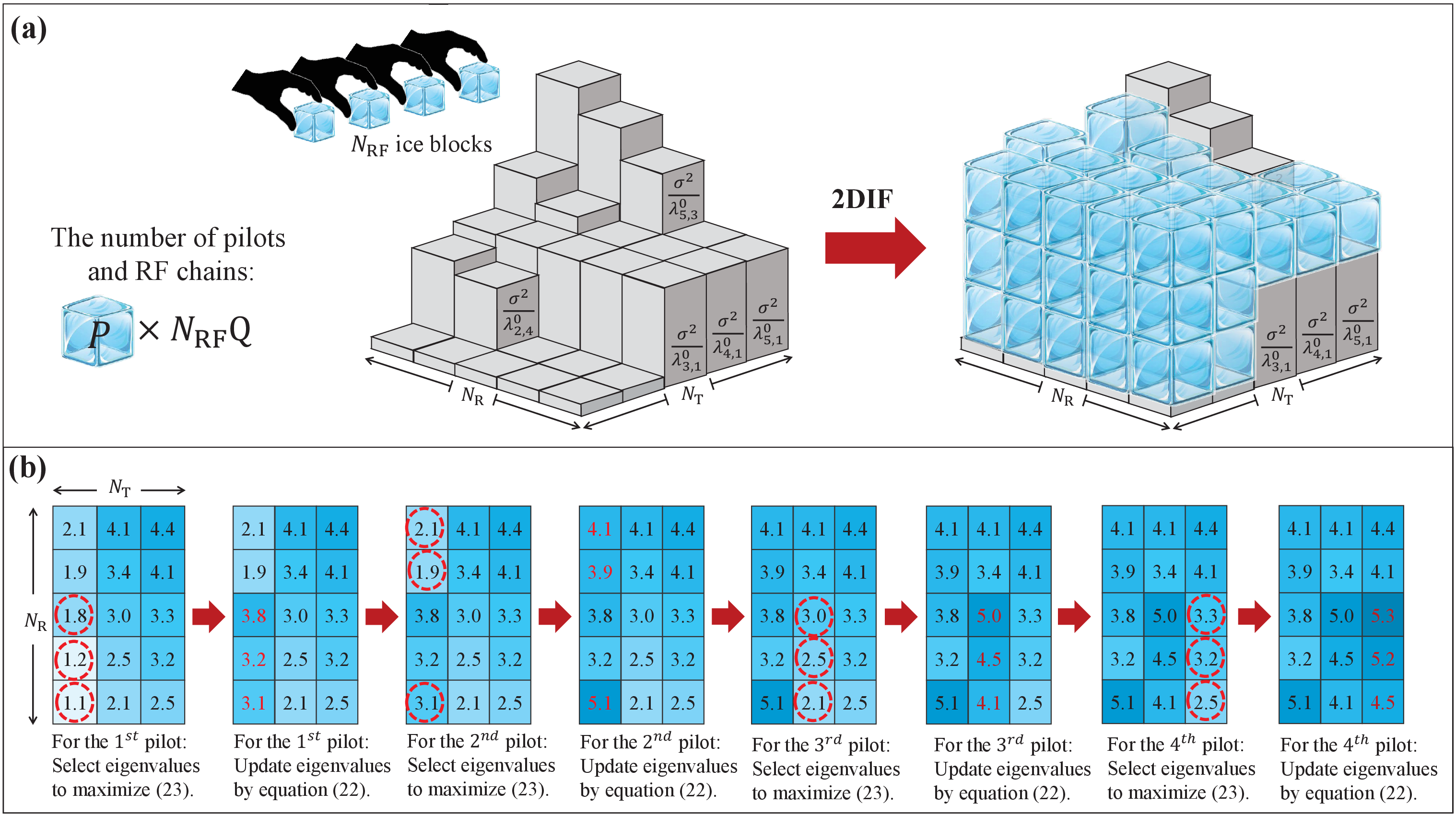}
	\caption{ (a) shows how the pilot allocation of the proposed 2DIF algorithm 
	works to maximize the MI. (b) provides an example to show the 
	implementation process of the 
	proposed 2DIF, where $N_{\rm R} = 5$, $N_{\rm T} = 3$, $N_{\rm RF} = 3$, 
	$P=2$, and $Q=4$. The number within the $({n_{\rm T}},n_{\rm R})$-th square 
	is the ice-level $\frac{{{\sigma ^2}}}{{{\lambda^t 
	_{{n_{\rm{T}}},{n_{\rm{R}}}}}}}$.
	}
	\label{img:GIF_vs_IF}
\end{figure*}
{\color{black}
	
In this subsection, we provide insightful explanations to the proposed 2DIF 
algorithm to clarify its physical significance. At first, the relationship between the well-known water-filling method and the proposed 2DIF method is discussed. Then, the superiority of the proposed 2DIF method over the existing IF method \cite{cui2024nearoptimal} is illustrated. 
	
\subsubsection{Ideal water-filling}	
To better understand the proposed observation matrix design, we first 
interpret problem (\ref{eq:Entropy}) from the view of \emph{water-filling}. 
Specifically, the orthogonal property of $\mb{W}_q$ proved in {\bf Lemma 
	\ref{lemma:Orth_W}} allows us to replace the noise 
covariance matrix ${\bm \Xi}$ in (\ref{eq:Entropy}) by $\sigma^2{\bf I}_{N_{\rm 
		RF}Q}$ without affecting the optimal ${I}({\bf y};{\bf h})$. 
Then, by further relaxing the constraints ${\bf W}_q^T{\bf W}_q = 
\mb{I}_{N_{\rm RF}}$ and $\|{\bf v}_q\|^2 = P$, we focus only on the total 
power 
constraint imposed on the overall observation matrix ${\bf X}$, i.e.,  
${\rm{Tr}}\left( {{{\bf{X}}}{\bf{X}}^H} \right) = 
P{N_{{\rm{RF}}}}Q$. In this case, the optimal value of problem 
(\ref{eq:Entropy}) is shown to have an upper bound: 
\begin{align}\label{eq:problem_t+1_relax}
	\notag
	\mathop {\max }\limits_{{\bf X}}~
	&\log_2\det \left( {{{\bf{I}}_{{N_{{\rm{RF}}}}Q}} + {1 \over {{\sigma 
	^2}}}{\bf{X}}^H{{\bf{\Sigma }}_{\bf h}}{{\bf{X}}}} \right) \\
	{\rm s.t.} ~& {\rm{Tr}}\left( {{{\bf{X}}}{\bf{X}}^H} 
	\right) =P {N_{{\rm{RF}}}}Q.
\end{align}
Notably, this upper bound is equivalent to the channel capacity of a 
point-to-point MIMO system equipped with  
$N_{\rm T}N_{\rm R}$ transmit antennas and $N_{\rm 
	RF}Q$ receive antennas. 
Thereafter, the overall observation matrix can be optimally solved 
as\footnote{For 
ease of 
discussion, we 
assume that $N_{\rm RF}Q$ is smaller than the rank of channel covariance, ${\bf 
\Sigma}_{\bf h}$.} 
\begin{align}\label{eq:opt_solution_wf}
	{{\bf{X}}}^{\rm ideal} = {{\bf{U}}_0}\left( {:,\left[ {1, \cdots 
			,{N_{{\rm{RF}}}Q}} \right]} \right){{\bf{P}}},  
\end{align}
where ${{\bf{P}}} = {\rm{diag}}\left( {\sqrt {{p_{1}}} , \cdots ,\sqrt 
{{p_{N_{{\rm{RF}}}Q}}} } \right)$ is the power allocation matrix. The power 
allocated to the $n$-th eigenvector is determined by the water-filling 
principle, i.e., ${p_{n}} = {\left({\beta  - {{{\sigma ^2}} \over {{\lambda 
_{n}}\left( {{{\bf{\Sigma }}_{\bf h}}} \right)}}} \right)^ + }$, where the 
water-level $\beta$ is adjusted to satisfy the total power constraint 
${\rm{Tr}}\left( {{{\bf{X}}^{\rm ideal}}({\bf{X}}^{\rm ideal})^H} \right) = \sum\nolimits_{n = 1}^{{N_{{\rm{RF}}}}Q} {{p_n}}  = P {N_{{\rm{RF}}}Q}$. 

Although the ideal observation matrix ${{\bf{X}}}^{\rm ideal}$, that maximizes 
the upper bound \eqref{eq:problem_t+1_relax}, might not be 
implementable in practice (as ${{\bf{X}}}^{\rm ideal}$ may violate the 
constraints ${\bf W}_q^T{\bf W}_q = \mb{I}_{N_{\rm RF}}$ and $\|{\bf v}_q\|^2 = 
P$), it can give us two pivotal intuitions. First, the observation matrix 
should 
align 
with the eigenspace $\{\mb{a}_{n_{\rm T}}\otimes \mb{b}_{n_{\rm R}}\}^{N_{\rm 
T},N_{\rm R}}_{n_{\rm T}=1,n_{\rm R}=1}$ of the 
full MIMO channel covariance, ${\bf \Sigma}_{\bf h} = {\bf \Sigma}_{\rm 
T}\otimes {\bf \Sigma}_{\rm R}$. 
Second, we need to fill in more power (water) to the eigenvectors having 
larger eigenvalues $\{\lambda_n({\bf \Sigma}_{\bf h})\}_{n=1}^{N_{\rm RF}Q}$ (or equivalently lower 
base 
levels $\{\frac{\sigma^2}{\lambda_n({\bf \Sigma}_{\bf h})}\}_{n=1}^{N_{\rm RF}Q}$). 

\subsubsection{2DIF versus water-filling}
The proposed 2DIF algorithm materializes the above two intuitions under the 
practical constraints  ${\bf W}_q^T{\bf W}_q = 
\mb{I}_{N_{\rm RF}}$ and $\|{\bf v}_q\|^2 = 
P$ via the eigenvector selection process in \eqref{eq:problem_t+1_reorg}. 
The first intuition is automatically 
achieved by 
assigning $\mb{v}_{t+1}$ with a eigenvector from $\{\sqrt{P}\mb{a}_{n_{\rm 
T}}^*\}_{n_{\rm T} = 
1}^{N_{\rm 
T}}$ and assigning the columns of $\mb{W}_{t+1}$ with different 
eigenvectors from 
$\{\mb{b}_{n_{\rm R}}\}_{n_{\rm R} = 1}^{N_{\rm 
R}}$, as proved in \textbf{Lemma~\ref{lemma:eigen_selection}}. The second 
intuition is approximately accomplished via selecting eigenvectors 
that have lower base levels, $\{\frac{\sigma^2}{\lambda_n({\bf \Sigma}_{\bf 
h})}\}$, by more times. This is attributed to the fact that the 
maximization of 
\eqref{eq:problem_t+1_reorg} tends to select an eigenvalue combination 
$\{\lambda^t_{n_{\rm T}, n_{{\rm R}, k}}\}_{k = 1}^{N_{\rm RF}}$ that has the 
lowest $\{\frac{\sigma^2}{\lambda^t_{n_{\rm T}, n_{{\rm R}, 
k}}}\}_{k = 1}^{N_{\rm RF}}$ on average. To see this approximation more clearly, 
we rewrite 
the updating rule of the selected eigenvalue in \eqref{eq:lambda_update} as
\begin{align}\label{eq:lambda_update_new}
	\underbrace {{{{\sigma ^2}} \over {\lambda _{{n_{\rm{T}}},{n_{\rm{R}}}}^{t 
	+ 1}}}}_{{\text{Updated ice-level}}} & = 
\underbrace {{{{\sigma ^2}} \over 
	{\lambda _{{n_{\rm{T}}},{n_{\rm{R}}}}^t}}}_{{\text{Current ice-level}}} + 
	\underbrace P_{{\text{Height of an ice block}}}.
\end{align}
		Equation \eqref{eq:lambda_update_new} 
		reveals that every time the eigenvector $ {{\bf{a}}_{{n_{\rm{T}}}}} 
		\otimes 
			{{\bf{b}}_{{n_{\rm{R}}}}}$ is selected, the value of ${{{{\sigma 
			^2}} \over 
					{\lambda _{{n_{\rm{T}}},{n_{\rm{R}}}}^t}}}$ increases by 
					$P$. 
Similar to the water-filling process, 
the process described by \eqref{eq:lambda_update_new} can be vividly 
interpreted as 
allocating an ice block having $P$-unit power to the 
$({n_{\rm{T}}},{n_{\rm{R}}})$-th 
orthogonal channel, where ${{{\sigma ^2}} \over 
	{\lambda_{{n_{\rm{T}}},{n_{\rm{R}}}}^t}}$ is viewed as the ice 
	level in the $t$-th timeslot. 
To summarize, due to the 
consideration of 
$N_{\rm RF}$ \ac{rf} chains and $N_{\rm T}\times N_{\rm R}$ MIMO systems, in 
each timeslot, the 2DIF first selects $N_{\rm RF}$ orthogonal channels, which 
have the 
deepest ice-levels $\{\frac{\sigma^2}{\lambda^t_{n_{\rm T}, n_{{\rm R}, 
			k}}}\}_{k = 1}^{N_{\rm RF}}$ on average, from the 
total $N_{\rm T}\times N_{\rm R}$ channels. Then, the 2DIF will fill $N_{\rm RF}$ ice 
blocks (i.e., 
$N_{\rm RF}$ pilots) of height $P$ onto them. 
As illustrated in Fig.~\ref{img:GIF_vs_IF} (a), after $Q$ timeslots, the final 
ice-levels of 
all channels can have a similar height with the water-level, $\beta$, determined by the water-filling principle. In this case, the second intuition is approximately 
achieved. 

\textbf{Example 2} (Example for Executing 2DIF). We present an example to show 
the growth of ice-level in Fig.~\ref{img:GIF_vs_IF} (b). Here, we set $N_{\rm 
R} = 
5$, $N_{\rm T} = 3$, $N_{\rm RF} = 3$, $P=2$, and $Q=4$.
The number table in Fig.~\ref{img:GIF_vs_IF} (b) records all ice-levels in each 
timeslot, i.e., $\frac{{{\sigma 
^2}}}{{{\lambda^t 
			_{{n_{\rm{T}}},{n_{\rm{R}}}}}}}$ for ${n_{\rm{T}}}\in\{1,2,3\}$ and 
			${n_{\rm{R}}}\in\{1,2,3,4,5\}$. For each pilot, we select $3$ 
deepest ice-levels, that maximize \eqref{eq:problem_t+1_reorg}, within one 
column of the number table, i.e., $\{1.1, 1.2, 1.8\}$ for the $1^{\rm st}$ 
pilot,  
$\{3.1, 1.9, 2.1\}$ for the $2^{\rm nd}$ pilot, $\{3.0, 2.5, 2.1\}$ for 
the $3^{\rm rd}$ pilot, and $\{2.5, 3.2, 3.3\}$ for 
the $4^{\rm th}$ pilot. Then, every selected ice-level is increased by $P = 2$ 
for eigenvalue update, i.e., ice-filling, and the corresponding eigenvectors are assigned to the 
precoders, $\mb{v}_t$, and combiners, $\mb{W}_t$. 
As a result, the ice-levels gradually grow and can finally approximate the 
ideal water-level, which is calculated as $\beta = 4.31$. 

}

	\subsubsection{2DIF versus IF}
	We now comprehensively compare the proposed 2DIF algorithm with the IF 
	algorithm devised in our prior work \cite{cui2024nearoptimal}.  
	In a nutshell, the IF algorithm is a special case of 2DIF algorithm, when 
	$N_{T} = 1$, $N_{\rm RF} = 1$, and $P = 1$. Specifically, the IF algorithm 
	is tailored to the channel estimation scenario of a	\emph{single-antenna} 
	transmitter with 
	\emph{unit} transmit power and  a multi-antenna receiver with \emph{single 
	RF 
	chain}. It only exploits the eigenvalue selection and updating rules of the 
	receive kernel $\boldsymbol{\Sigma}_{\rm R}$ to design combiners. 
	Attributed to the general design in this paper, our proposed 2DIF has the following two 
	advantages over IF. 
	\begin{itemize}
		\item \textbf{Utilizing the correlation at transmitter}: The proposed 2DIF 
		algorithm reveals the 
		eigenvalue/eigenspace evolution mode of the full MIMO channel kernel, 
		$\boldsymbol{\Sigma}_{\rm T} \otimes \boldsymbol{\Sigma}_{\rm R}$. It  
		enables the 
		joint precoder and combiner design to simultaneously exploit both the transmit and 
		receive channel covariance. The IF algorithm, however, lacks the 
		exploitation of the channel covariance at the transmitter.
		Thereby, the proposed 2DIF algorithm enjoys a better channel estimation performance. 
		\item \textbf{Utilizing multi-RF-chain observations}: The IF algorithm is not 
		suitable to multi-RF-chain receivers. Specifically, IF algorithm 
		can only produce one combiner vector $\mb{w}_t \in \mathbb{C}^{N_{\rm 
		R}\times 1}$ in each timeslot $t$ via the eigenvector-selection 
		process. Thus, it is possible that the combiners $\mb{w}_t$ and 
		$\mb{w}_{t+1}$ designed in two adjacent timeslots are selected as 
		the  
		same eigenvector. At this moment, if $\mb{w}_t$ and $\mb{w}_{t+1}$ are used 
		in the same timeslot $q$ in a multi-RF-chain 
		receiver, they will lead to duplicate observations: 
		$\mb{w}_t^H(\mb{H}\mb{v}_q + \mb{z}_q) = \mb{w}_{t+1}^H(\mb{H}\mb{v}_q 
		+ \mb{z}_q)$.
		These  make no 
		contribution to the channel estimation accuracy.  
		In contrast, the proposed 2DIF algorithm efficiently circumvents this 
		issue by  
		simultaneously producing  
		all $N_{\rm RF}$ combiner vectors, i.e., $\mb{W}_t \in 
		\mathbb{C}^{N_{\rm 
		R}\times N_{\rm RF}}$, in each timeslot. The orthogonal constraint $\mb{W}_t^H\mb{W}_t = \mb{I}$ 
		ensures that the columns of  $\mb{W}_t$ are selected as different 
		eigenvectors $\left\{ {\bf{b}}_{{n_{{\rm R}, k}}} \right\}_{k 
		= 
		1}^{{N_{\rm{R}}}}$. This fact guarantees the discrepancy of the observed channel information by each RF chain, contributing to improved channel
		estimation accuracy. 
	\end{itemize}

\section{Proposed Two-Stage 2DIF (TS-2DIF) Based Observation Matrix Design}\label{sec:TS-2DIF}
Turn now to the receiver architecture with phase-only controllable 
analog combiner presented in Fig. 
\ref{img:system} (b). As the coefficients of matrices 
$\{\mb{W}_q = \mb{A}_q\mb{D}_q\}_{q = 1}^Q$ can no longer be freely 
manipulated, the 
proposed 2DIF 
algorithm might not be 
implementable in these scenarios. 
To address this problem, a TS-2DIF algorithm is proposed in 
this section.


\subsection{Overview of TS-2DIF Observation Matrix Design}

Recall that the phase-only 
controllable structure in Fig.~\ref{img:system} (b) requires to express the 
hybrid combiner as
$\mb{W}_q = \mb{A}_q\mb{D}_q$. Each element of the analog combiner  
$\mb{A}_q$ is restricted by the modulus constraint $|\mb{A}_q(n_{\rm R}, n_{\rm 
rf})| = \frac{1}{\sqrt{N_{\rm 
R}}}$, for $n_{\rm R}\in\{1,\cdots,N_{\rm R}\}$ and $n_{\rm 
rf}\in\{1,\cdots,N_{\rm 
RF}\}$. This poses a structural constraint on 
the feasible set of 
$\mb{W}_q$, destroying its eigenvector structure, thereby making the 2DIF 
algorithm inapplicable.  
Therefore, it becomes necessary to redesign a new set of 
observation matrices $\{\mb{v}_q\}_{q = 1}^{Q}$ and $\{\mb{W_q} = 
\mb{A}_q\mb{D}_q\}_{q = 1}^{Q}$ tailored for the phase-only controllable 
architecture. 


\begin{algorithm}[!t]
	\caption{TS-2DIF Based Combiner and Precoder Design} 
	\begin{algorithmic}[1]\label{alg:twostage}
		\REQUIRE  
		Number of pilots $Q$, kernel ${\bm \Sigma}_{\bf h}$.
		\ENSURE 
		Designed precoders $\{\mb{v}^{\rm opt}_q\}_{q=1}^Q$ and hybrid combiners 
		$\{\mb{A}^{\rm opt}_q\}_{q=1}^Q$ and $\{\mb{D}^{\rm opt}_q\}_{q=1}^Q$. \\
		\emph{Stage 1 (Optimal observation matrix design) }		 
		\STATE Obtain the ideal precoders $\{\mb{v}_q^{\rm IF}\}_{q = 
			1}^Q$ and overall 
		combiners  $\{\mb{W}^{\rm IF}_q\}_{q = 1}^Q$ from 
		\textbf{Algorithm~\ref{alg:2DIF}}
		\STATE Get the ideal observation matrix $\mb{X}_q^{\rm IF} = 
		({\mb{v}_q^{\rm 
				IF}})^*\otimes \mb{W}_q^{\rm IF}$ for all $q\in\{1,\cdots,Q\}$ \\
		\emph{Stage 2 (Joint hybrid combiner and precoder design)}	
		\FOR{$q = 1, \cdots, Q$}
		\WHILE{no convergence of $\|\mb{X}_q^{\rm IF} -  
			\mb{v}_q^*\otimes 
			(\mb{A}_q\mb{D}_q)\|_F^2$}
		\STATE Update the digital combiner $\mb{D}_q$ by \eqref{eq:Dopt}
		\STATE Update the analog combiner $\mb{A}_q$ by \eqref{eq:Aopt}
		\STATE Update the precoder $\mb{v}_q$ by \eqref{eq:vopt}
		\ENDWHILE
		\ENDFOR
		\RETURN Designed precoders $\{\mb{v}_q\}_{q=1}^Q$ and hybrid combiners 
		$\{\mb{A}_q\}_{q=1}^Q$ and $\{\mb{D}_q\}_{q=1}^Q$ for channel estimation.
	\end{algorithmic}
\end{algorithm}

For this purpose, we propose a TS-2DIF algorithm as summarized in 
\textbf{Algorithm~3}, which consists of two stages. 
In the first stage, \textbf{Algorithm~1} is carried out to obtain 
the ideal observation matrix $\mb{X}_q^{\rm IF} = ({\mb{v}_q^{\rm 
IF}})^*\otimes \mb{W}_q^{\rm IF}$ for $\forall q$, where 
 the superscript ``IF" is used to indicate the 
observation matrices generated by the 2DIF algorithm.  Subsequently, the 
second stage aims to 
make the newly designed observation 
matrix  $\mb{X}_q = \mb{v}_q^*\otimes (\mb{A}_q\mb{D}_q)$ sufficiently close to 
the ideal observation matrix $\mb{X}_q^{\rm IF}$. To this end, the joint 
optimization of $\{\mb{v}_q\}_{q = 1}^{Q}$ and $\{\mb{A}_q\mb{D}_q\}_{q = 
1}^{Q}$ 
are formulated as:
%
\begin{align}
	\min_{\mb{v}_q, \mb{A}_q, \mb{D}_q} &\:\: \|\mb{X}_q^{\rm IF} -  
	\mb{v}_q^*\otimes 
	(\mb{A}_q\mb{D}_q)\|_F^2, \label{eq:P1} \\
	\mathrm{s.t.}\:\:& \:\: \|\mb{v}_q\|^2 = P,
	\tag{\ref{eq:P1}a} \label{eq:P1a} \\
	& \:\: |{{\bf{A}}_q}| = \frac{1}{{\sqrt {{N_{\rm{R}}}} }}{{\bf{1}}_{{N_{\rm{R}}} \times {N_{{\rm{RF}}}}}},
	\tag{\ref{eq:P1}b} \label{eq:P1b}
\end{align}
where ${\bf{1}}_{{N_{\rm{R}}} \times {N_{{\rm{RF}}}}}$ is an ${N_{\rm{R}}}$-by-${N_{{\rm{RF}}}}$ all-one matrix.
By solving problem \eqref{eq:P1}, the newly designed observation 
matrices $\{\mb{X}_q\}_{q=1}^Q$ are expected to achieve a comparable channel 
estimation performance with  $\{\mb{X}_q^{\rm IF}\}_{q=1}^Q$.

\subsection{Joint Optimization of Precoders and Hybrid Combiners}
It is intricate to directly solve problem \eqref{eq:P1}
owing to the 
non-convex modulus constraint in \eqref{eq:P1b} as well as the coupled 
relationship of 
$\mb{v}_q$, $\mb{A}_q$, and $\mb{D}_q$ in the objective \eqref{eq:P1}.
To overcome these challenges, we exploit the alternating minimization method 
to iteratively update $\mb{A}_q$ and $\mb{D}_q$, and $\mb{v}_q$ until an 
convergence condition 
triggers. 
The detailed optimization procedures are elaborated one by one as follows.

\subsubsection{Fix $\mb{A}_q$ and $\mb{v}_q$, and optimize $\mb{D}_q$} 
For ease of discussion, we denote $\mb{v}_q = [v_q(1), v_q(2), \cdots, 
v_q(N_{\rm T})]^T$ and define 
$\mb{X}_{q, n}^{\rm IF} \in \mathbb{C}^{N_{\rm R}\times N_{\rm 
		RF}}$ as the $n$-th block component of $\mb{X}_q^{\rm IF}$ such that 
		$\mb{X}_q^{\rm IF} = 
[(\mb{X}_{q, 
	1}^{\rm IF})^T, (\mb{X}_{q, 2}^{\rm IF})^T,\cdots, (\mb{X}_{q, N_{\rm 
	T}}^{\rm 
	IF})^T]^T$.
Then, when keeping the combiner $\mb{A}_q$ and the precoder 
$\mb{v}_q$ fixed, 
the sub-problem for optimizing $\mb{D}_q$ is expressed as 
\begin{align}\label{eq:P1.2}
\min_{\mb{D}_q} \sum_{n = 1}^{N_{\rm T}}\|\mb{X}_{q, 
	n}^{\rm IF} - 
{v}_q^*(n)\mb{A}_q\mb{D}_q \|_F^2.
\end{align}
Problem \eqref{eq:P1.2} is a standard quadradic programming (QP), which can be 
optimally solved by making the gradient of objective function to zero, i.e.,
\begin{align}\label{eq:Dopt}
	\mb{D}_q^{\rm opt} &= \left(\sum_{n = 1}^{N_{\rm 
	T}}|v_q(n)|^2\mb{A}_q^H\mb{A}_q\right)^{-1}\left(\sum_{n = 1}^{N_{\rm 
	T}}v_q(n)\mb{A}_q^H\mb{X}_{q,n}^{\rm IF}\right) \notag \\
&\overset{(a)}{=} \sum_{n = 1}^{N_{\rm 
		T}}\frac{v_q(n)}{P}\left(\mb{A}_q^H\mb{A}_q\right)^{-1}\mb{A}_q^H\mb{X}_{q,n}^{\rm
		 IF},
\end{align}
where (a) holds because $\|\mb{v}_q\|^2 = \sum_{n = 1}^{N_T} |v_q(n)|^2 = P$. 

\subsubsection{Fix $\mb{D}_q$ and $\mb{v}_q$, and optimize $\mb{A}_q$} We then 
fix the digital combiner $\mb{D}_q$ and precoder $\mb{v}_q$, and seeks an 
analog combiner that optimizes the following sub-problem:
\begin{align}
		\min_{\mb{A}_q} &\:\: \sum_{n = 1}^{N_{\rm T}}\|\mb{X}_{q, 
			n}^{\rm IF} - 
		\mb{A}_q\mb{D}_q {v}_q^*(n)\|_F^2, \label{eq:P1.3} \\
	\mathrm{s.t.}
	& \:\: |{{\bf{A}}_q}| = \frac{1}{{\sqrt {{N_{\rm{R}}}} }}{{\bf{1}}_{{N_{\rm{R}}} \times {N_{{\rm{RF}}}}}}.
	\tag{\ref{eq:P1.3}a} \label{eq:P1.3a}
\end{align}
Directly optimizing problem \eqref{eq:P1.3} is challenging owing to the 
constant 
modulus 
constraint \eqref{eq:P1.3a} and the product of $\mb{A}_q$ 
and $\mb{D}_q$. 
To address this issue, we notice that the objective function has the an 
upper bound due to the Cauchy-Schwarz inequality,
\begin{align}\label{eq:bound1}
	 &\sum_{n = 1}^{N_{\rm T}}\|\mb{X}_{q, 
		n}^{\rm IF} - 
	\mb{A}_q\mb{D}_q {v}_q^*(n)\|_F^2 \notag \\&\quad\quad{\le} 
	\sum_{n = 
	1}^{N_{\rm T}}\|\mb{X}_{q, 
		n}^{\rm IF}\mb{D}_q^{-1} - 
	\mb{A}_q {v}_q^*(n)\|_F^2 \| \mb{D}_q\|_F^2. 
\end{align}
In \eqref{eq:bound1}, the analog combiner $\mb{A}_q$ has escaped from the 
product form with $\mb{D}_q$, which can significantly simplify the 
optimization problem. Taking this into account, we replace the original 
objective function with $\sum_{n = 
	1}^{N_{\rm T}}\|\mb{X}_{q, 
	n}^{\rm IF}\mb{D}_q^{-1} - 
\mb{A}_q {v}_q^*(n)\|_F^2$.   
Then, by defining $\mb{J}_{q,n} = \mb{X}_{q, n}^{\rm IF}\mb{D}_q^{-1}$, the new 
objective function can be further simplified as {
\begin{align}\label{eq:bound2}
	&\sum_{n = 
		1}^{N_{\rm T}}\|\mb{J}_{q,n} - 
	\mb{A}_q {v}_q^*(n)\|_F^2 \notag \\
	& = C_1 +  \sum_{n = 
		1}^{N_{\rm T}}\left(\|\mb{A}_q\|_F^2\| |v_q(n)|^2 - 
	2\mathrm{Tr}\left\{\Re\left(v_q^*(n)\mb{J}_{q,n}^H\mb{A}_q\right)\right\}\right)\notag\\
	& = C_1 + N_{\rm RF}P - 
	2\mathrm{Tr}\left(\Re\left\{\mb{J}^H\mb{A}_q\right\}\right),
\end{align}}
where $C_1 = \sum_{n = 1}^{N_{\rm T}}\|\mb{J}_{q,n}\|_F^2$ and $\mb{J}:=\sum_{n 
= 
	1}^{N_{\rm T}}v_q(n)\mb{J}_{q,n}$. By 
combining \eqref{eq:bound1} and 
\eqref{eq:bound2}, the new optimization problem is formulated as 
\begin{align}
		\max_{|\mb{A}_q| = \frac{1}{\sqrt{N_{\rm R}}}{{\bf{1}}_{{N_{\rm{R}}} 
		\times 
		{N_{{\rm{RF}}}}}}}
		&\:\: 
		\mathrm{Tr}\left(\Re\left\{\mb{J}^H\mb{A}_q\right\}\right), 
			\label{eq:P1.4} 
\end{align}
Evidently, the optimal solution of \eqref{eq:P1.4} is given by 
\begin{align}\label{eq:Aopt}
\mb{A}_q^{\rm opt} = \frac{1}{\sqrt{N_{\rm 
R}}}\exp\left(j\angle\mb{J}\right),
\end{align}
which completes the update of $\mb{A}_q$ in step 6 of 
\textbf{Algorithm~\ref{alg:twostage}}. 
%
%

\subsubsection{Fix $\mb{A}_q$ and $\mb{D}_q$, and optimize $\mb{v}_q$} With 
given combiner matrices $\mb{A}_q$ and $\mb{D}_q$, 
the objective function in 
\eqref{eq:P1} for optimizing $\mb{v}_q$ can be rewritten as
\begin{align}\label{eq:vq}
 &\sum_{n = 1}^{N_{\rm 
	T}}\|\mb{X}_{q, 
		n}^{\rm IF} - 
	\mb{A}_q\mb{D}_q {v}_q^*(n)\|_F^2 \notag\\
	&\quad\quad\quad{=} \|\mb{X}_{q}^{\rm IF}\|_F^2 + 
	\|\mb{A}_q\mb{D}_q\|_F^2 - 
	2\Re\{\mb{c}_q^H \mb{v}_q\},
\end{align}
where $\mb{c}_q = [\mathrm{Tr}\{(\mb{X}_{q, 1}^{\rm IF})^H\mb{A}_q\mb{D}_q\}, 
\cdots, 
\mathrm{Tr}\{(\mb{X}_{q, N_T}^{\rm IF})^H\mb{A}_q\mb{D}_q\}]^T$. 
By further considering the power constraint in \eqref{eq:P1a}, the 
optimal $\mb{v}_q^{\rm opt}$ is given as\footnote{This work assumes the digital precoder $\mb{v}_q$ because it is technically feasible and commercially relevant for user  equipment. In the analog precoding case, (\ref{eq:vopt}) should be modified as ${\bf{v}}_q^{{\rm{opt}}} = \sqrt {P/{N_{\rm{T}}}} \exp \left( {j\angle {{\bf{c}}_q}} \right)$. In the hybrid precoding case, as proved in \cite{zhang2014achieving}, vector ${\bf{v}}_q^{{\rm{opt}}}$ in (\ref{eq:vopt}) can be perfectly realized by the hybrid precoder with at least 2 RF chains. 
}
\begin{align}\label{eq:vopt}
	\mb{v}_q^{\rm opt} = \sqrt{P}{\mb{c}_q}/{\|\mb{c}_q\|}. 
\end{align}
This completes the update of $\mb{v}_q$ in step 7 of 
\textbf{Algorithm~\ref{alg:twostage}}. 

To summarize, {\it Stage 2} of 
\textbf{Algorithm~\ref{alg:twostage}} alternatively updates $\mb{D}_q$, $\mb{A}_q$, and 
$\mb{v}_q$ using \eqref{eq:Dopt}, \eqref{eq:Aopt}, and \eqref{eq:vopt} until 
convergence. 
After obtaining the hybrid combiners and precoders for pilot 
transmission, the channel estimator \eqref{eq:postmean} can be 
utilized to  recover wireless channel matrices. 

\section{Computational Complexity Analysis and Kernel Selection}\label{sec:KS}
\subsection{Computational Complexity Analysis}

Consider {\bf Algorithm \ref{alg:2DIF}} at first. The EVDs 
for ${\bf \Sigma}_{\rm T}$ and ${\bf \Sigma}_{\rm R}$ require ${\cal 
O}\left(N^3_{\rm T}+N^3_{\rm R}\right)$ FLOPS. The update of eigenvalues 
requires ${\cal O}\left(QN_{\rm RF}\right)$ FLOPS in total. In {\bf Algorithm 
\ref{alg:TS_eig}},  the linear search requires the sort operations with the 
complexity of ${\cal O}\left(N_{\rm T}N_{\rm R}\log_2(N_{\rm R})\right)$, and 
calculating the objective requires ${\cal O}\left(N_{\rm T}N_{\rm RF}\right)$ 
FLOPS. Thus, the overall complexity of {\bf Algorithm \ref{alg:2DIF}} is ${\cal 
O}\left( {N_{\rm{T}}^3 + N_{\rm{R}}^3 + {N_{\rm{T}}}{N_{\rm{R}}}{{\log 
}_2}({N_{\rm{R}}}) + \left( {Q + {N_{\rm{T}}}} \right){N_{{\rm{RF}}}}} 
\right)$. The computational complexity of {\bf Algorithm \ref{alg:twostage}} is 
dominated by the alternating optimizations of $\mb{D}_q$, $\mb{A}_q$, and 
$\mb{v}_q$. In particular, their computations require ${\cal O}(Q( 
N_{{\rm{RF}}}^2{N_{\rm{R}}}  + N_{{\rm{RF}}}^3 ))$, ${\cal O}(Q( 
N_{{\rm{RF}}}^2{N_{\rm{R}}} + N_{{\rm{RF}}}^3 + 
{N_{\rm{T}}}{N_{\rm{R}}}{N^2_{{\rm{RF}}}} ))$, and ${\cal O}( 
{Q{N_{\rm{T}}}{N_{\rm{R}}}N_{{\rm{RF}}}^2})$ FLOPS, respectively. Assuming that 
the number of iterations is $I_o$, the overall computational complexity of {\bf 
Algorithm \ref{alg:twostage}} is $O\left( I_o{Q\left( { N_{{\rm{RF}}}^3 + 
{N_{\rm{T}}}{N_{\rm{R}}}N_{{\rm{RF}}}^2} \right)} \right)$.

It is worth noting that, {\bf Algorithms 1}$\sim${\bf 3} only rely on 
the given kernel ${\bf \Sigma}_{\bf h}$ instead of the instantaneous channels 
or received pilots, thus they do not need to be implemented in real time. Since 
the channel covariance does not change so frequently, the designed observation 
matrices can be deployed online for channel estimation for a long time. 
Thereby, from the long-term perspective, the computational complexity of online 
deploying the proposed channel estimator is not dominated by the 
observation matrix design. This advantage allows the proposed 2DIF to achieve lower computational complexity compared with the existing channel estimators \cite{vlachos2023time,Vlachos2019MIMO}.

\subsection{Kernel Selection}\label{subsec:kernel_sel}
Selecting an appropriate covariance matrix, i.e., kernel $\bm \Sigma$, is crucial for constructing a robust estimator. The kernel $\bm \Sigma$ dictates the shape and adaptability of the estimator, thereby influencing its performances to detect functional trends and provide precise predictions. Given the localized-correlation characteristic of MIMO channels, the ideal kernel should enhance the similarity between adjacent antennas while diminishing its impact as the distance increases. In this section, two kinds of kernels are recommended.
\subsubsection{Statistical Kernel}
Given the kernel's role as the prior covariance of channel $\bf h = {\rm 
vec}(\mb{H})$, 
the optimal strategy is to utilize the actual covariance for channel 
estimation, i.e., ${\bm \Sigma}_{\bf h}={\mathsf E}\left({\rm 
	vec}(\mb{H}){\rm 
	vec}(\mb{H})^{\rm 
H}\right)$. Prior to deploying the proposed estimator online, it is 
feasible to train an approximation of ${\bm \Sigma}_{\bf h}$ in advance by 
leveraging some existing channel models or channel datasets 
\cite{werner2009estimating,Sungwoo'18,Karthik'18,Khalilsarai;20}. Concretely, 
according to the law of large numbers, ${\bm \Sigma}_{\bf h} $ can be trained 
by 
\begin{eqnarray}
{\bm \Sigma}_{\bf h} \approx \frac{1}{R}\sum\limits_{r = 1}^R {{\rm 
		vec}(\mb{H}_r){\rm 
		vec}(\mb{H}_r)^{\rm 
		H}},
\end{eqnarray}
where $R$ is the number of channel realizations and ${\bf{H}}_r$ is the $r$-th 
channel realization used for kernel training. As for the covariance matrices 
$\mb{\Sigma}_{\rm T}$ and  $\mb{\Sigma}_{\rm R}$ that characterize the transmitter-side 
and receive-side channel covariance, respectively, they can be obtained by 
${{\bf{\Sigma }}_{\rm{T}}} = 
\frac{1}{{{RN_{\rm{R}}}}}\sum\nolimits_{r = 1}^{R}\sum\nolimits_{n = 
1}^{{N_{\rm{R}}}}{ 
{{{ {{\bf{H}}_r^T\left( {n,:} \right)} }}{{\bf{H}}_r^*}\left( 
{n,:} 
\right)} }$ and ${{\bf{\Sigma }}_{\rm{R}}} = 
\frac{1}{{{RN_{\rm{T}}}}}\sum\nolimits_{r = 1}^{R}\sum\nolimits_{n = 
1}^{{N_{\rm{T}}}}  
{{\bf{H}}_r\left( {:,n} \right){{ {{\bf{H}}^H_r\left( {:,n} \right)} 
}}} $.

\subsubsection{Artificial Kernels}
In practical scenarios where obtaining an explicit channel model or channel dataset is challenging, it is preferred to train an artificial kernel to replace ${\bf \Sigma}_{\bf h}$ \cite{Jieao'ICC'24}. For simplicity, we assume that the uniform linear arrays (ULAs) are deployed at both the \ac{bs} and the user, while it can be easily extended to the UPA case. The consistency between the artificial kernels and the statistical kernel is that both of them assign higher similarity to nearby antennas and decrease influence with distance. Given array parameters, the mutual coupling matrices ${\bf C}_{\rm rx}$ and ${\bf C}_{\rm tx}$ can be calculated or measured. Thus here we focus on characterizing the spatial correlations ${\bf R}_{\rm rx}$ and ${\bf R}_{\rm tx}$, two artificial kernels are recommended \cite{srinivas2012information}:
		
		 {\it i) Laplace kernel}: The Laplace kernel ${\bm \Sigma}_{\rm La} $ is the most popular choice in Bayesian estimation. Let ${{\bf{n}}_{\rm T}} = 
		\left[ { - \frac{{{N_{\rm T}} - 1}}{2}, - \frac{{{N_{\rm T}} - 3}}{2}, \cdots 
			,\frac{{{N_{\rm T}} - 1}}{2}} \right]^T$ and ${{\bf{n}}_{\rm R}} = \left[ { - 
			\frac{{{N_{\rm R}} - 1}}{2}, - \frac{{{N_{\rm R}} - 3}}{2}, \cdots ,\frac{{{N_{\rm R}} - 1}}{2}} \right]^T$.  Then, the Laplace kernels, which respectively characterize the spatial correlations at the user and the \ac{bs}, can be modeled as
	\begin{align}
	{{\bf{R }}_{{\rm La},{\rm T}}} =  \exp \left( { - {\eta 
			^2}{\frac{d^2}{\lambda^2}}{{\left| {{{\bf{1}}^T_{{N_{\rm T}}}} 
					\otimes 
					{{\bf{n}}_{\rm T}} - {\bf{n}}_{\rm T}^T \otimes 
					{{\bf{1}}_{{N_{\rm T}}}}} \right|}^{ 
				\odot 2}}} \right),\\
	{{\bf{R }}_{{\rm La},{\rm R}}} = \exp \left( { - {\eta 
			^2}{\frac{d^2}{\lambda^2}}{{\left| {{{\bf{1}}^T_{{N_{\rm R}}}} 
					\otimes 
					{{\bf{n}}_{\rm R}} - {\bf{n}}_{\rm R}^T \otimes 
					{{\bf{1}}_{{N_{\rm R}}}}} \right|}^{ 
				\odot 2}}} \right),
\end{align}
where $\eta>0$ is an adjustable hyperparameter; and ${\bf Z}^{\odot 2}$ 
denotes the element-wise product of two matrices $\bf Z$. Thus, the overall 
kernel can be written 
as ${{\bf{\Sigma }}_{{\rm{La}}}} = ( {{{({\bf{C}}_{{\rm{tx}}}^{1/2})}^T}{{\bf{R}}_{{\rm{La}},{\rm{T}}}}{{({\bf{C}}_{{\rm{tx}}}^{1/2})}^*}} ) \otimes ( {{\bf{C}}_{{\rm{rx}}}^{1/2}{{\bf{\Sigma }}_{{\rm{La}},{\rm{R}}}}{{({\bf{C}}_{{\rm{rx}}}^{1/2})}^H}} )$. 

 {\it ii) Bessel kernel:} By exploiting the inherent periodic property of Bessel functions, the Bessel kernel, denoted as ${\bm \Sigma}_{\rm Be}$, is well-suited for recovering data with oscillatory patterns. The Bessel kernels, which respectively characterizes the spatial correlations at the user and the \ac{bs}, can be modeled as
\begin{align}
	{{\bf{R }}_{{\rm{Be}},{\rm T}}}  = {J_0}\left( {\eta 
		\frac{d}{\lambda}\left| {{{\bf{1}}^T_{{N_{\rm T}}}} 
			\otimes 
			{{\bf{n}}_{\rm T}} - {\bf{n}}_{\rm T}^T \otimes 
			{{\bf{1}}_{{N_{\rm T}}}}} \right|} \right),\\
	{{\bf{R }}_{{\rm{Be}},{\rm R}}}  = {J_0}\left( {\eta 
		\frac{ d}{\lambda}\left| {{{\bf{1}}^T_{{N_{\rm R}}}} 
			\otimes 
			{{\bf{n}}_{\rm R}} - {\bf{n}}_{\rm R}^T \otimes 
			{{\bf{1}}_{{N_{\rm R}}}}} \right|} \right), 
\end{align}
where $J_0$ is the zero-order Bessel function of the first kind and $\eta>0$ is 
a hyperparameter. The 
overall kernel is written as $ {{\bf{\Sigma }}_{{\rm{Be}}}} = ( {{{({\bf{C}}_{{\rm{tx}}}^{1/2})}^T}{{\bf{R}}_{{\rm{Be}},{\rm{T}}}}{{({\bf{C}}_{{\rm{tx}}}^{1/2})}^*}} ) \otimes ( {{\bf{C}}_{{\rm{rx}}}^{1/2}{{\bf{\Sigma }}_{{\rm{Be}},{\rm{R}}}}{{({\bf{C}}_{{\rm{rx}}}^{1/2})}^H}} )$.

The hyperparameter $\eta$ plays a pivotal role in closely mirroring the real channel covariance, whose value can be determined by a \ac{ml} estimator. We assume that $R$ channel realizations are utilized to train a kernel $\bf \Sigma$, where $\bf \Sigma$ can be either ${\bm \Sigma}_{\rm La}$ or ${\bm \Sigma}_{\rm Be}$. Then, the estimation of $\eta$ is written as
	\begin{align}\label{eq:eta_opt}
		\eta^{\rm opt} = \mathop{\arg\!\max}_{\eta>0}~~{\sum\limits_{r = 1}^R}\ln\left({\mathsf P}\left({\bf y}_r|\eta\right)\right), 
	\end{align}
	wherein the likelihood function is given by
	\begin{align}
		{\mathsf P}\left({\bf y}_r|\eta\right) = {{\exp \left( { - {\bf{y}}_r^{{H}}{{\left( {\bf{X}}_r^{{H}}{\bf{\Sigma }}{{\bf{X}}_r} + {\bf{\Xi}}_r \right)}^{ - 1}}{{\bf{y}}_r}} \right)} \over {{\pi ^{QN_{\rm RF}}}\det \left( {{\bf{X}}_r^{{H}}{\bf{\Sigma }}{{\bf{X}}_r} + {\bf{\Xi}}_r } \right)}},
	\end{align}
	in which ${\bf y}_r = {\bf X}_r^{H}{\bf h}_r + {\bf z}_r \in{\mathbb C}^{QN_{\rm RF}}$ denotes the received pilot associated with the $r$-th channel realization ${{\bf{h}}_r}$ for kernel training; ${{\bf{X}}_r}$ is obtained based on the randomly generated precoders and combiners; and ${\bf{\Xi}}_r$ is the covariance of \ac{awgn} ${\bf z}_r$. One-dimensional search method is adopted to obtain $\eta^{\rm opt}$.

\subsubsection{Adaptive Kernel}
In cases where the statistical kernel is unavailable and the artificial kernels are inaccurate, we introduce a novel adaptive kernel training strategy. It uses channels estimated by the 2DIF algorithm to adaptively update the channel kernel for implementing 2DIF.  The method smoothly integrates the kernel training and channel estimation stages, eliminating the additional time period required for estimating the covariance matrix. 

\begin{figure*}[b]
	\hrulefill	
	\begin{align}\label{eq:flow}
		\hat{\mb \Sigma }_T^{(0)} \otimes \hat{\mb \Sigma }_R^{(0)} \underbrace{\overset{\rm 2DIF}{\rightarrow} \hat{\mb{H}}_{1} \overset{\eqref{eq:kernel_learning}}{\rightarrow} 
			\hat{\mb \Sigma }_T^{(1)} \otimes \hat{\mb \Sigma }_R^{(1)}}_{\rm frame\:1} 
		\underbrace{\overset{\rm 2DIF}{\rightarrow} \hat{\mb{H}}_{2} \overset{\eqref{eq:kernel_learning}}{\rightarrow}  
			\hat{\mb \Sigma }_T^{(2)} \otimes \hat{\mb \Sigma }_R^{(2)}}_{\rm frame\:2} \overset{\rm 2DIF}{\rightarrow} 
		\cdots 
		\underbrace{\overset{\rm 2DIF}{\rightarrow} 
			\hat{\mb{H}}_{T_f} \overset{\eqref{eq:kernel_learning}}{\rightarrow}  \hat{\mb \Sigma }_T^{(T_f)} \otimes \hat{\mb \Sigma }_R^{(T_f)}}_{{\rm frame}\:T_f}. 
	\end{align}
\end{figure*}

Consider $T_f$ consecutive frames, where the inter-frame channels $\mb{H}_{t_f}\in\mathbb{C}^{N_R\times N_T}$, $\forall {t_f}\in\{1,2,\cdots,T_f\}$ are i.i.d distributed, following ${\rm vec}(\mb{H}_{t_f})\sim \mathcal{CN}(0, {\bf{\Sigma }}_{\rm T} \otimes  {\bf{\Sigma }}_{\rm R})$. Our strategy iteratively estimates the channels $\hat{\mb{H}}_{t_f}$ and updates the kernels $\hat{\mb \Sigma }_T^{(t_f)}$ and $\hat{\mb \Sigma }_R^{(t_f)}$, targeting at gradually improving the accuracy of  $\hat{\mb{H}}_{t_f}$ and converging $\hat{\mb \Sigma }_T^{(t_f)} \otimes \hat{\mb \Sigma }_R^{(t_f)}$ to the real kernel ${\bf{\Sigma }}_{\rm T} \otimes  {\bf{\Sigma }}_{\rm R}$. To be specific, at frame $t_f$, the channel $\mb{H}_{t_f}$ is estimated by 2DIF using the kernel $\hat{\mb \Sigma }_T^{(t_f - 1)} \otimes \hat{\mb \Sigma }_R^{(t_f - 1)}$ trained during frames $1 \sim t_f - 1$:
\begin{align}
	\hat{\mb \Sigma }_T^{(t_f - 1)} \otimes \hat{\mb \Sigma }_R^{(t_f - 1)} \overset{\rm 2DIF}{\rightarrow} \hat{\mb{H}}_{t_f}. 
\end{align}
Then, we accumulate the information of the newly estimated channel $\hat{\mb{H}}_{t_f}$ to update the kernels $\hat{\mb \Sigma }_T^{(t_f)}$ and $\hat{\mb \Sigma }_R^{(t_f)}$ as
\begin{align}\label{eq:kernel_learning}
	\left\{
	\begin{array}{c}
		\hat{\mb \Sigma }_T^{(t_f)} = \frac{t_f - 1}{t_f}\hat{\mb \Sigma }_T^{(t_f - 1)} + \frac{1}{t_fN_R}\sum_{n = 1}^{N_R} \hat{\mb{H}}_{t_f}^T(n, :)\hat{\mb{H}}_{t_f}^*(n, :)  \\
		\hat{\mb \Sigma }_R^{(t_f)} = \frac{t_f - 1}{t_f}\hat{\mb \Sigma }_R^{(t_f - 1)} + \frac{1}{t_fN_T}\sum_{n = 1}^{N_T} \hat{\mb{H}}_{t_f}(:, n)\hat{\mb{H}}_{t_f}^H(:, n)
	\end{array}
	 \right..
\end{align}
As a result, the proposed adaptive kernel training strategy proceeds as in \eqref{eq:flow}. 
To trigger the working flow in \eqref{eq:flow}, the initial kernels $\hat{\mb \Sigma }_T^{(0)}$ and $\hat{\mb \Sigma }_R^{(0)}$ are set as identity matrices: $\hat{\mb \Sigma }_T^{(0)} = {\bf I}_{N_T}$ and $\hat{\mb \Sigma }_R^{(0)} = {\bf I}_{N_R}$. 

The proposed adaptive kernel learning method exhibits two key advantages. Firstly, as the initial kernels are identity matrices, the proposed method requires no prior information about the real kernels, making it applicable to general and practical communication systems. Secondly, the proposed method allows the coexistence of channel estimation and kernel training within one frame, as it leverages the channels estimated by 2DIF itself to train the kernel. This integration eliminates the additional time period required for kernel learning, greatly simplifying the frame structure and  protocol of 2DIF in practical systems.

\section{Simulation Results}\label{sec:sim}
\subsection{Simulation Setup and Baselines}
We consider a single-user MIMO system, where ULAs are equipped on both the \ac{bs} and the user. The general correlated Rayleigh-fading channel model in (\ref{eq:channel_model}) is considered to generate $\bf H$. In specific, the carrier frequency is set to 3.5 GHz.
Following the setup in \cite{Pizzo'22'TWC}, the elements of the spatial correlation matrices ${\bf R}_{\rm rx}$ are calculated via (\ref{eqn:R_rx_m_n}), and the spatial correlation matrices ${\bf R}_{\rm tx}$ is obtained from the similar process. As a typical example to analyze antenna coupling effect \cite{MathWorks5GToolbox}, here we consider the dipole antennas with $\lambda/2$ length and $\lambda/100$ width for both transceivers. The antenna spacing is set to be ${\lambda}/{8}$. In this context, the spatial-scattering function $f_{\rm rx}(\varphi, \theta)$ for the BS and that for the user $f_{\rm tx}(\varphi, \theta)$ are given by $f_{\rm rx}(\varphi, \theta)=f_{\rm tx}(\varphi, \theta)=\frac{1.67}{2 \pi} \cos ^4(\theta)$ \cite{balanis2016antenna}. Given these array parameters, the mutual coupling matrices ${\bf C}_{\rm rx}$ and ${\bf C}_{\rm tx}$ are calculated using Matlab Antenna Toolbox \cite{MathWorks5GToolbox}. Otherwise specifically
specified, the numbers of transceiver antennas and \ac{rf} chains are set as: $N_{\rm T}=4$, $N_{\rm R}=64$, and $N_{\rm RF}=4$, respectively. The \ac{snr} is defined as ${\rm SNR} = \frac{P}{\sigma^2} {\mathsf E}(\|{\bf h}\|^2)$, whose default value is set to 10 dB. The evaluation criterion of estimation accuracy is the \ac{nmse}, which is expressed as ${\rm NMSE} = {\mathsf E}(\frac{\|{\bf h}-{\hat{\bf h}}\|^2}{\|{\bf h}\|^2})$. The number of channel realizations for kernel training is set to $R=100$. The default value of pilot length is set to $Q=48$.

To verify the effectiveness of the proposed 2DIF based channel estimator and TS-2DIF based channel estimator, the following seven schemes are simulated for comparison:
\begin{itemize}
	\item {\bf LS}: The LS channel estimator is feasible only when observation dimension is no less than the channel dimension, i.e., $QN_{\rm RF}\geq N_{\rm T}N_{\rm R}$. To realize this, we assume the pilot length is $Q = \left\lceil {{{{N_{\rm{T}}}{N_{\rm{R}}}} / {{N_{{\rm{RF}}}}}}} \right\rceil=64$, and all combiners/precoders are generated by discrete Fourier transform (DFT) matrices.
	\item {\bf MMSE}: Under the same setting of LS estimator, the classic MMSE estimator with DFT observation matrices is implemented to recover channel $\bf 
	H$ via (\ref{eq:postmean}).
	\item {\bf AMP}: Utilizing the channel sparsity in angular domain, the approximate message passing (AMP) method proposed in \cite{rangan2019vector} is implemented to estimate channel $\bf H$. The combiners and precoders are randomly generated from Gaussian random measurement matrices.
	\item {\bf IF scheme}: By viewing the considered MIMO system as $N_{\rm T}$ 
	independent SIMO systems, the IF-based channel estimator 
	\cite{cui2024nearoptimal} can be utilized to recover $\bf H$ in a 
	column-by-column way. Note that, since IF scheme is only applicable to the 
	single-RF-chain case, the pilot length used should be modified as 
	$\left\lceil 
	{QN_{\rm RF}} \right\rceil=192$ to 
	ensure 
	that it has the same number of observations with the 2DIF.  
	\item {\bf Proposed 2DIF}: The proposed 2DIF method in {\bf Algorithm \ref{alg:2DIF}} is employed to design the precoders and combiners of \ac{mimo} system in Fig. \ref{img:system} (a). Based on these designed observation matrices/vectors, the MMSE estimator in  (\ref{eq:postmean}) is employed to estimate channel $\bf H$.
	\item {\bf Proposed TS-2DIF}: The proposed TS-2DIF method in {\bf Algorithm \ref{alg:twostage}} is employed to design the precoders and combiners of \ac{mimo} system in Fig. \ref{img:system} (b). The MMSE estimator in  (\ref{eq:postmean}) is employed to recover channel $\bf H$.
	\item {\bf Ideal water-filling}: To provide a fundamental performance limit, the ideal (but may not be practically achievable) observation matrices $\{{\bf X}_q\}_{q=1}^Q$ are directly obtained by solving (\ref{eq:problem_t+1_relax}) via water-filling method. Then, (\ref{eq:postmean}) is employed to recover channel $\bf H$.
\end{itemize}
\begin{figure}[!t]
	\centering
	\includegraphics[width=0.5\textwidth]{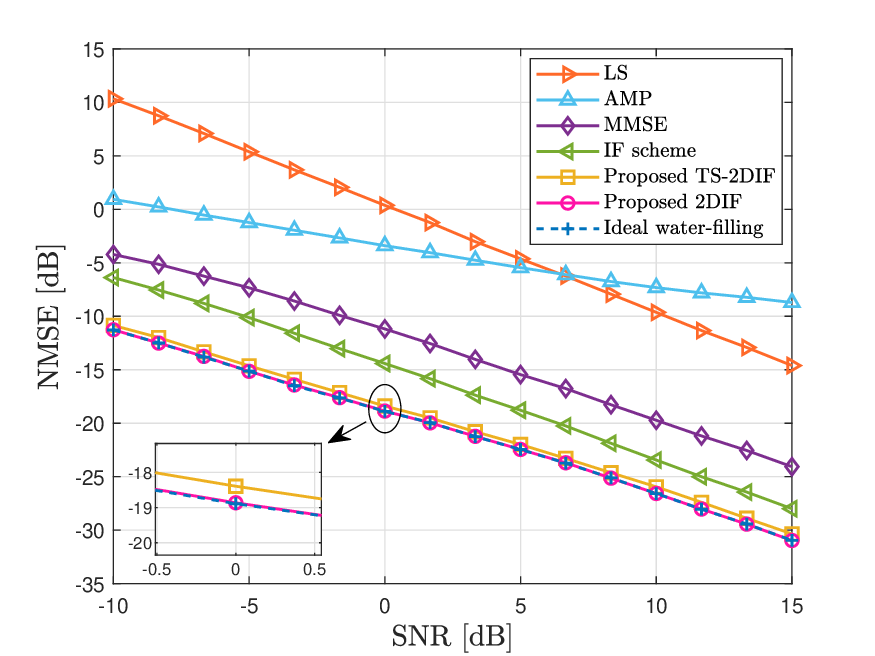}
	\caption{The NMSE as a function of SNR for different schemes.
	}
	\label{img:sim_NMSE_SNR}
\end{figure}

\subsection{Estimation Accuracy Under Statistical Kernel}
In this subsection, we consider the ideal case when the statistical kernel ${\bf \Sigma}_{\bf h}:={\mathsf E}({\bf h}{\bf h}^H)$ can be trained thanks to the known channel models or datasets. Then, ${\bf \Sigma}_{\bf h}$ is employed for all required estimators for the channel recovery.

Firstly, we plot the NMSE as a function of SNR in Fig.~\ref{img:sim_NMSE_SNR}. 
One can observe that, thanks to the carefully designed observation 
matrices/vectors, the proposed 2DIF and TS-2DIF schemes remarkably outperform 
the benchmark schemes in estimation accuracy. The reason is that the proposed 
methods fully exploit the spatial correlations among the transceiver antennas 
for channel estimation. In particular, the NMSEs for the proposed 2DIF and 
TS-2DIF schemes are about 5 dB lower than that for the IF scheme. It is because 
the IF schemes realizes the MIMO channel estimation by viewing it as $N_{\rm 
T}$ independent SIMO channel estimations, which ignores the spatial 
correlation of transmitter antennas. Besides, we note that the proposed 2DIF 
scheme achieves very similar performance to the ideal water-filling scheme. 
This phenomenon implies that the practical pilot allocation can behave almost 
the same as the theoretically-optimal ``continuous'' pilot allocation.

\begin{figure}[!t]
	\centering
	\includegraphics[width=0.5\textwidth]{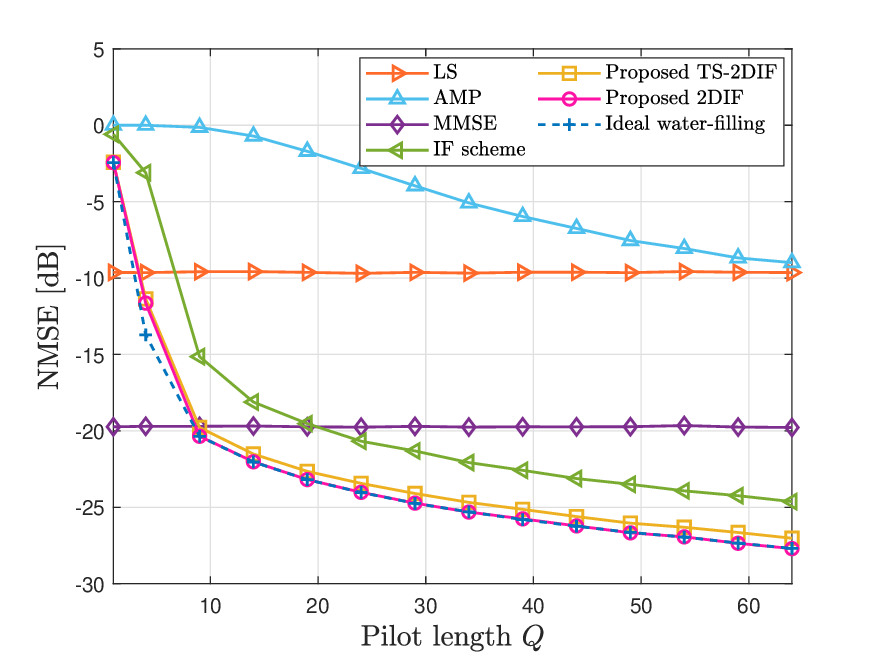}
	\caption{The NMSE as a function of pilot length $Q$ for different schemes.
	}
	\label{img:sim_NMSE_Q}
\end{figure}

\begin{figure}[!t]
	\centering
	\includegraphics[width=0.5\textwidth]{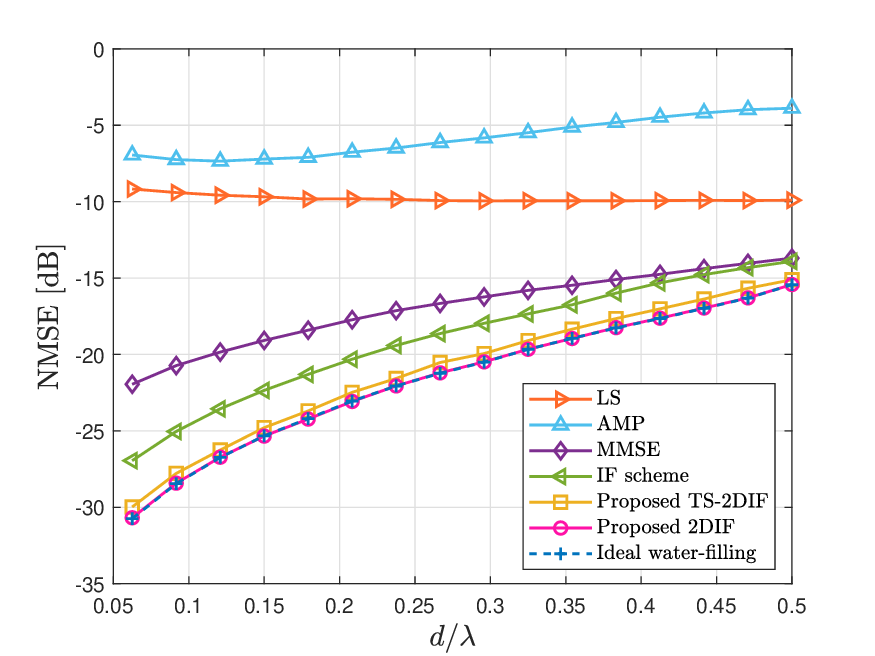}
	\caption{The NMSE as a function of the normalized antenna spacing $d/\lambda$ for different schemes.
	}
	\label{img:sim_NMSE_d}
\end{figure}

Then, the NMSE versus the number of pilots $Q$ is provided in Fig. \ref{img:sim_NMSE_Q}. One can find that the superiority of the proposed schemes still holds. Although the dimension of the estimated parameters is high as $N_{\rm T}N_{\rm R}=256$, using a small number of pilots $Q=20$, the NMSEs for the proposed schemes can be lower than $-20$ dB. In contrast, even if the pilot length is longer than $Q=60$, the conventional AMP estimator is still unable to achieve such high accuracy. It indicates that utilizing the correlation of compact antennas is of great significance for high-accuracy channel estimation. In addition, observing Fig. \ref{img:sim_NMSE_SNR} and Fig. \ref{img:sim_NMSE_Q}, one can conclude that the TS-2DIF scheme can achieve almost the same estimation accuracy as the 2DIF. This indicates that, from the perspective of \ac{csi} acquisition, both hybrid \ac{mimo} structures in Fig. \ref{img:system} have no obvious performance gap.

To observe the impact of spatial correlations on the estimation accuracy, we plot the NMSE as a function of the normalized antenna spacing $d/\lambda$ in Fig. \ref{img:sim_NMSE_d}. One can observe that, as the antenna spacing decreases, the estimation accuracy of the proposed schemes becomes higher. It is because a smaller antenna spacing leads to stronger spatial correlations, which can provide more prior knowledge for Bayesian estimators. In this case, the more informative kernel allows the proposed schemes to realize more accurate channel estimation.  As the antenna spacing increases, due to the reduced channel correlation, the proposed schemes gradually converge to the classical MMSE scheme.   However, even if the antenna spacing is $\lambda/2$, the proposed 2DIF and TS-2DIF methods can still hold the superiority. This fact indicates that, for a conventional massive \ac{mimo} system, as long as its channels are not i.i.d. Rayleigh-fading (otherwise ${\bf \Sigma}_{\bf h}={\bf I}_{N_{\rm T}N_{\rm R}}$), the non-diagonal kernel ${\bf \Sigma}_{\bf h}$ with some structural properties can still contribute to the improvement of the estimation accuracy. 

\subsection{Estimation Accuracy Under Artificial Kernels}
\begin{figure}[!t]
	\centering
	\includegraphics[width=0.5\textwidth]{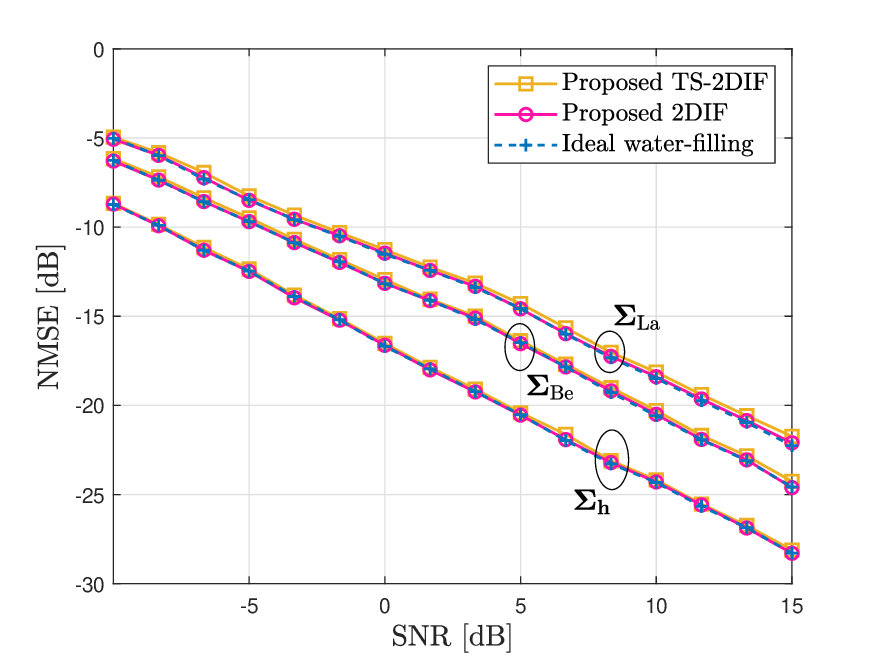}
	\caption{The NMSE as a function of SNR for different kernels.
	}
	\label{img:sim_NMSE_SNR_art}
\end{figure}
\begin{figure}[!t]
	\centering
	\includegraphics[width=0.5\textwidth]{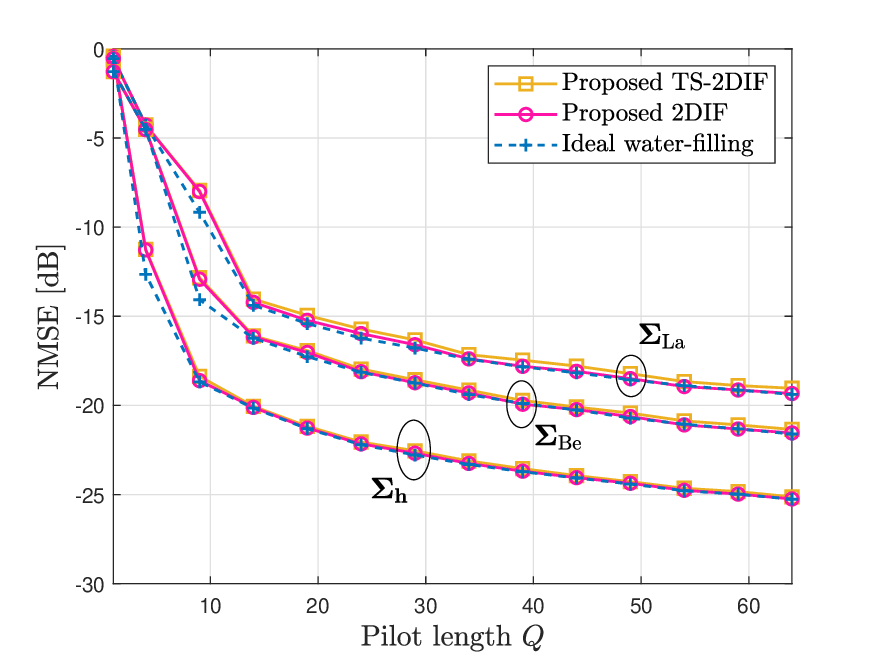}
	\caption{The NMSE as a function of pilot length $Q$ for different kernels.
	}
\label{img:sim_NMSE_Q_art}
\end{figure}

In practical scenarios where obtaining an explicit channel model or channel dataset is challenging, it is preferred to use an artificial kernel for channel estimation, as discussed in Subsection \ref{subsec:kernel_sel}. In this subsection, two popular artificial kernels, i.e., Laplace kernel ${\bf \Sigma}_{\rm La}$ and Bessel kernel ${\bf \Sigma}_{\rm Be}$, are compared with the ideal statistical kernel ${\bf \Sigma}_{\bf h}$. We plot the NMSE as a function of the SNR in Fig. \ref{img:sim_NMSE_SNR_art} and the NMSE as a function of the pilot length $Q$ in Fig. \ref{img:sim_NMSE_Q_art}, respectively.

One can observe that, for each type of kernel, the three proposed schemes exhibit very similar trends in estimation accuracy. This implies that our proposed estimators have the similar robustness for different kernels in channel estimation. Compared to the ideal kernel, the performance losses for both artificial kernels are acceptable. For examples, when ${\rm SNR}=10$ dB, the NMSEs for Laplace kernel, Bessel kernel, and statistical kernel are about -18 dB, -20 dB, and -24 dB, respectively. When the pilot length is $Q=19$, the NMSEs for these three kernels are about -15 dB, -17 dB, and -21 dB, respectively. We can conclude that, even if the real covariance (statistical kernel) is unknown, the proposed channel estimator can still hold their performance advantages by training artificial kernels. This fact encourages the potential applications of the proposed schemes in practice.


\subsection{Estimation Accuracy Under Adaptive Kernel Training}
\begin{figure}[!t]
	\hspace{-1.2em}
	\centering
	\includegraphics[width=0.5\textwidth]{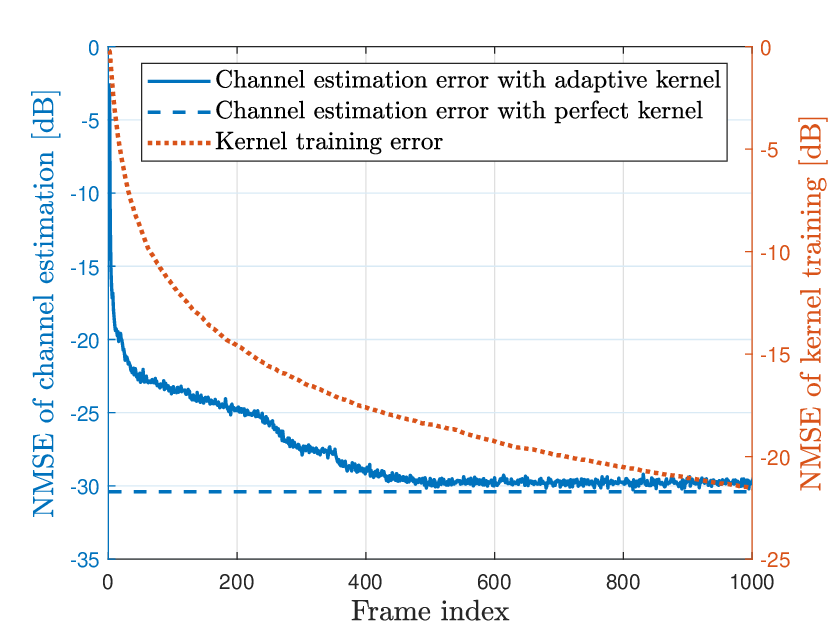}
	\caption{The NMSE performance of the adaptive kernel training strategy.
	}
	\label{img:Timeslot_identity}
\end{figure}

We now evaluate the performance of the proposed adaptive kernel training strategy. Fig.~\ref{img:Timeslot_identity} illustrates the NMSE of the estimated channel $\hat{{\bf H}}_{t_f}$ by 2DIF, accompanied by the NMSE of learned kernel $\hat{\mb \Sigma }_T^{(t_f)} \otimes \hat{\mb \Sigma }_R^{(t_f)}$ by adaptive training. The frame index $t_f$ increases from 0 to 1000. The other settings are as follows: $N_T = 4$, $N_R = 64$, $d = \frac{\lambda}{8}$, $Q = 48$, ${\rm SNR} = 15\:{\rm dB}$. It is observed that, as the frame index increases, the proposed 2DIF algorithm and adaptive kernel training method achieve a mutual beneficial relationship. The newly estimated channels continuously improve the fidelity of learned kernels, while a more accurate kernel further decreases channel estimation error in future frames. Particularly, the NMSE of channel estimation rapidly declines below -20 dB after 16 frames of kernel training and closely approaches that achieved by the perfect kernel after 400 frames. Moreover, as the channel covariance matrix almost remains unchanged over time, we can keep using the learned kernel after adaptive training to achieve a near-optimal channel estimation performance (see frames 500$\sim$1000). This fact validates the feasibility and superiority of the proposed designs in practical situations where no prior information about the kernel is available. 

\section{Conclusions}\label{sec:con}
By fully exploiting the channel correlations among antennas, this work has proposed a generalized channel estimation framework for densifying MIMO systems, focusing on the design of observation matrices. By maximizing the MI between channels and received pilots, the 2DIF method has been proposed to design observation matrices through jointly optimizing the precoders and combiners. Subsequently, the TS-2DIF method has been proposed to extend the applicability of our framework to the typical hybrid MIMO whose analog combiner is phase-only controllable. Simulation results have validated the superiority of our proposed channel estimation schemes. 


\appendices

\section{Proof of Lemma \ref{lemma:Sigma_decomp}}\label{appendix:Sigma_decomp}

	Given the channel model in (\ref{eq:channel_model}) and ${\bf h} \equiv {\rm vec}\left({\bf H}\right)$, the vectorized channel can be rewritten as 
	\begin{align}
		\notag
		{\bf{h}} & = {\rm{vec}}\left( {{\bf{C}}_{{\rm{rx}}}^{1/2}{\bf{R}}_{{\rm{rx}}}^{1/2}{{\bf{H}}_{{\rm{iid}}}}{\bf{R}}_{{\rm{tx}}}^{1/2}{\bf{C}}_{{\rm{tx}}}^{1/2}} \right)\\
		& = \left( {{{\left( {{\bf{R}}_{{\rm{tx}}}^{1/2}{\bf{C}}_{{\rm{tx}}}^{1/2}} \right)}^T} \otimes \;\left( {{\bf{C}}_{{\rm{rx}}}^{1/2}{\bf{R}}_{{\rm{rx}}}^{1/2}} \right)} \right){\rm{ vec}}\left( {{{\bf{H}}_{{\rm{iid}}}}} \right),
	\end{align}
	where the second equation holds since ${\mathop{\rm vec}\nolimits} \left( {{\bf{ABC}}} \right) = \left( {{{\bf{C}}^T} \otimes {\bf{A}}} \right){\rm{vec}}\left( {\bf{B}} \right)$.
	Then, the covariance of channel $\bf h$ can be derived as (\ref{eq:Sigma_h_derive}), where $(a)$ holds since ${\mathsf E}\left({{\rm{vec}}\left( {{{\bf{H}}_{{\rm{iid}}}}} \right){{\left( {{\rm{vec}}\left( {{{\bf{H}}_{{\rm{iid}}}}} \right)} \right)}^H}}\right)={\bf I}_{N_{\rm R}N_{\rm T}}$, $(b)$ holds due to the commutative law of Kronecker product $\left( {{\bf{A}} \otimes 
		{\bf{B}}} \right)\left( {{\bf{C}} \otimes {\bf{D}}} \right) = \left( 
	{\bf{AC}}\right) \otimes \left({\bf{BD}} \right)$; and $(c)$ holds by defining 
	\begin{align}
		{{\bf{\Sigma }}_{\rm{T}}} & = {{{\left( {{\bf{C}}_{{\rm{tx}}}^{1/2}} \right)}^T}{\bf{R}}_{{\rm{tx}}}^*{{\left( {{\bf{C}}_{{\rm{tx}}}^{1/2}} \right)}^*}},	 \\
		{{\bf{\Sigma }}_{\rm{R}}} & ={{\bf{C}}_{{\rm{rx}}}^{1/2}{\bf{R}}_{{\rm{rx}}}{{\left( {{\bf{C}}_{{\rm{rx}}}^{1/2}} \right)}^H}}.
	\end{align}
	One can find that, the matrix ${{\bf{\Sigma 
		}}_{\rm{T}}}$ only depends on the spatial correlation matrix ${\bf R}_{\rm tx}$ and the mutual coupling matrix ${\bf C}_{\rm tx}$ at the user, while the matrix ${{\bf{\Sigma }}_{\rm{R}}}$ is 
	only associated with the spatial correlation matrix ${\bf R}_{\rm rx}$ and the mutual coupling matrix ${\bf C}_{\rm rx}$ at the BS. Thus, ${\bf{\Sigma }}_{\rm T}$ and ${\bf{\Sigma }}_{\rm R}$ can be 
	viewed as the kernels that characterize the correlation among the transmitter 
	antennas and that among the receiver antennas, respectively. 
	This completes the proof.
	\begin{figure*}[b]
		\hrulefill	
		\begin{align}\label{eq:Sigma_h_derive}
			\notag
			{{\bf{\Sigma }}_{\bf{h}}} =  {\mathsf E}\left( {{\bf{h}}{{\bf{h}}^H}} 
			\right) & = {\mathsf E}\left( {\left( {{{\left( {{\bf{R}}_{{\rm{tx}}}^{1/2}{\bf{C}}_{{\rm{tx}}}^{1/2}} \right)}^T} \otimes \;\left( {{\bf{C}}_{{\rm{rx}}}^{1/2}{\bf{R}}_{{\rm{rx}}}^{1/2}} \right)} \right){\rm{ vec}}\left( {{{\bf{H}}_{{\rm{iid}}}}} \right){{\left( {{\rm{vec}}\left( {{{\bf{H}}_{{\rm{iid}}}}} \right)} \right)}^H}\left( {{{\left( {{\bf{R}}_{{\rm{tx}}}^{1/2}{\bf{C}}_{{\rm{tx}}}^{1/2}} \right)}^*} \otimes \;{{\left( {{\bf{C}}_{{\rm{rx}}}^{1/2}{\bf{R}}_{{\rm{rx}}}^{1/2}} \right)}^H}} \right)} \right)  \\
			&  \stackrel{(a)}{=}
			{\left( {{{\left( {{\bf{R}}_{{\rm{tx}}}^{1/2}{\bf{C}}_{{\rm{tx}}}^{1/2}} \right)}^T} \otimes \;\left( {{\bf{C}}_{{\rm{rx}}}^{1/2}{\bf{R}}_{{\rm{rx}}}^{1/2}} \right)} \right){\rm{ }}\left( {{{\left( {{\bf{R}}_{{\rm{tx}}}^{1/2}{\bf{C}}_{{\rm{tx}}}^{1/2}} \right)}^*} \otimes \;{{\left( {{\bf{C}}_{{\rm{rx}}}^{1/2}{\bf{R}}_{{\rm{rx}}}^{1/2}} \right)}^H}} \right)} 
			\notag \\ & \stackrel{(b)}{=}  {\left( {{{\left( {{\bf{C}}_{{\rm{tx}}}^{1/2}} \right)}^T}{\bf{R}}_{{\rm{tx}}}^*{{\left( {{\bf{C}}_{{\rm{tx}}}^{1/2}} \right)}^*}} \right) \otimes \left( {{\bf{C}}_{{\rm{rx}}}^{1/2}{\bf{R}}_{{\rm{rx}}}{{\left( {{\bf{C}}_{{\rm{rx}}}^{1/2}} \right)}^H}} \right)}
			\stackrel{(c)}{=} {{\bf{\Sigma }}_{\rm{T}}} \otimes {{\bf{\Sigma }}_{\rm{R}}}.
		\end{align}
	\end{figure*} 

\begin{figure*}[b]
	\hrulefill	
	\begin{align}\label{eq:Orth_W_derivation}
		\notag
		I({{\bf{y}}};{\bf{h}})  &\stackrel{(a)}{=} {\log _2}\det  {\bigg (} 
		{{\bf{I}}_{{N_{\rm{R}}}{N_{\rm{T}}}}} + \\ \notag & ~~~~~{1 \over {{\sigma 
					^2}}}\left[ {{\bf{v}}_1^* \otimes {{\bf{W}}_1}, \cdots ,{\bf{v}}_Q^* 
			\otimes {{\bf{W}}_Q}} \right]{\rm{blkdiag}}\left( {{{\left( 
					{{\bf{W}}_1^H{{\bf{W}}_1}} \right)}^{ - 1}}, \cdots ,{{\left( 
					{{\bf{W}}_Q^H{{\bf{W}}_Q}} \right)}^{ - 1}}} \right)\left[ {{\bf{v}}_1^* 
			\otimes {\bf{W}}_1, \cdots ,{\bf{v}}_Q^* \otimes {\bf{W}}_Q} 
		\right]^H{{\bf{\Sigma }}_{\bf{h}}} {\bigg )}  \\ 
		&  \stackrel{(b)}{=} {\log_2}\det \left( 
		{{{\bf{I}}_{{N_{\rm{R}}}{N_{\rm{T}}}}} + \frac{1}{{{\sigma 
						^2}}}\sum_{q = 1}^Q \left(\left( {{\bf{v}}_q^*{\bf{v}}_q^T} \right) 
			\otimes 
			\left( {{{\bf{W}}_q}{{\left( {{\bf{W}}_q^H{{\bf{W}}_q}} \right)}^{ - 
						1}}{\bf{W}}_q^H} \right)\right){{\bf{\Sigma }}_{\bf{h}}}} \right).
	\end{align}
\end{figure*} 

\section{Proof of Lemma \ref{lemma:Orth_W}}\label{appendix:Orth_W}
Using some matrix techniques, the MI $I({{\bf{y}}};{\bf{h}})$ can be rewritten as equation (\ref{eq:Orth_W_derivation}), where $(a)$ holds since $\det \left( {{\bf{I}} + {\bf{AB}}} \right) = \det \left( {{\bf{I}} + {\bf{BA}}} \right)$ and ${\bm \Xi} = \sigma^2{\rm blkdiag}\left( 
{ {\bf W}_1^H{\bf W}_1, \cdots ,{\bf W}_Q^H{\bf W}_Q} \right)$; $(b)$ holds 
according to the property that $\left( {{\bf{a}} \otimes {\bf{B}}} \right) 
{\bf{C}} \left( {{{\bf{a}}}^H \otimes {\bf{D}}} \right) = \left( 
{{{\bf{a}}{\bf{a}}^H}} 
\right) \otimes \left( {{\bf{BCD}}} \right)$ if all dimensions meet the 
requirements of matrix multiplications. To find more insights, we perform 
singular value decomposition (SVD) on all $\{{\bf W}_q\}_{q=1}^Q$ and then 
substitute all decomposition formulas ${{\bf{W}}_q} = 
{{\bf{\Pi}}_q}{{\bf{\Omega}}_q}{\bf{\Upsilon}}_q^H$ into 
(\ref{eq:Orth_W_derivation}). It is evident that ${{\bf{W}}_q}{\left( 
	{{\bf{W}}_q^H{{\bf{W}}_q}} \right)^{ - 1}}{\bf{W}}_q^H = 
{{\bf{\Pi}}_q}{\bf{\Pi}}_q^H$, thus the MI $I({{\bf{y}}};{\bf{h}})$ can be 
rewritten as
\begin{align}\label{eq:Orth_W_derivation_c}
	\notag
	&I({{\bf{y}}};{\bf{h}}) =  \\ & {\log _2}\det \left( {{{\bf{I}}_{{N_{\rm{R}}}{N_{\rm{T}}}}} + \frac{1}{{{\sigma ^2}}}\sum\limits_{q = 1}^Q {\left( {\left( {{\bf{v}}_q^*{\bf{v}}_q^T} \right) \otimes \left( {{{\bf{\Pi }}_q}{\bf{\Pi }}_q^H} \right)} \right)} {{\bf{\Sigma }}_{\bf{h}}}} \right).
\end{align}
Observing (\ref{eq:Orth_W_derivation_c}), one can find that the MI $I({{\bf{y}}};{\bf{h}})$ in (\ref{eq:Entropy}) only relies on the orthogonal matrix ${\bf \Pi}_q\in{\mathbb C}^{N\times N_{\rm RF}}$ decomposed from ${\bf W}_q$ for all $q\in\{1,\cdots,Q\}$, while it does not depend on any ${{\bf{\Omega}}_q}$ or ${\bf{\Upsilon}}_q$. It indicates that imposing ${\bf W}_q={\bf \Pi}_q$ does not change the value of $I({{\bf{y}}};{\bf{h}})$. As a result, the orthogonality constraint ${\bf{W}}_q^H{{\bf{W}}_q} = {\bf{\Pi}}_q^H{{\bf{\Pi}}_q} = {{\bf{I}}_{{\rm{RF}}}}$ can be safely introduced into the problem formulation regarding $I({{\bf{y}}};{\bf{h}})$, which completes the proof.

\section{Proof of MI increment $I({\bf{\bar y}}_{t+1};\mb{h}) - 
	I({\bf{\bar y}}_t; \mb{h})$}\label{appendix:MI_incre}
Using some matrix partition operations, the MI $I({\bf{\bar y}}_{t+1};\mb{h})$ can be rewritten as
\begin{align}
	\notag
	&I({{{\bf{\bar y}}}_{t + 1}};{\bf{h}}) \stackrel{(a)}{=} {\log _2}\det \left( {{{\bf{I}}_{{N_{{\rm{RF}}}}Q}} + {1 \over {{\sigma ^2}}}{\bf{\bar X}}_{t + 1}^H{{\bf{\Sigma }}_{\bf{h}}}{{{\bf{\bar X}}}_{t + 1}}} \right) \\
	&= \log_2\det \left[
	\begin{array}{cc}
		\!\!{{{\bf{I}}_{{N_{{\rm{RF}}}}t}} \!+\! {1 \over {{\sigma ^2}}}{\bf{\bar X}}_t^H{{\bf{\Sigma }}_{\bf{h}}}{{{\bf{\bar X}}}_t}}\!\!\!\!\!\! & 
		{{1 \over {{\sigma ^2}}}{\bf{\bar X}}_t^H{{\bf{\Sigma }}_{\bf{h}}}{{\bf{X}}_{t + 1}}} \\
		{{1 \over {{\sigma ^2}}}{\bf{X}}_{t + 1}^H{{\bf{\Sigma }}_{\bf{h}}}{{{\bf{\bar X}}}_t}} & 
		{{{\bf{I}}_{{N_{{\rm{RF}}}}}} \!+\! {1 \over {{\sigma ^2}}}{\bf{X}}_{t + 1}^H{{\bf{\Sigma }}_{\bf{h}}}{{\bf{X}}_{t + 1}}} \!\!
	\end{array}
	\right]
	\notag 
	\\
	&\stackrel{(b)}{=} \log_2\det \left[
	\begin{array}{cc}
		\!\!{{{\bf{I}}_{{N_{{\rm{RF}}}}t}} \!+\! {1 \over {{\sigma ^2}}}{\bf{\bar X}}_t^H{{\bf{\Sigma }}_{\bf{h}}}{{{\bf{\bar X}}}_t}}\!\!\!\!\!\! & 
		{{1 \over {{\sigma ^2}}}{\bf{\bar X}}_t^H{{\bf{\Sigma }}_{\bf{h}}}{{\bf{X}}_{t + 1}}} \\
		{{{\bf{0}}_{{N_{{\rm{RF}}}} \times {N_{{\rm{RF}}}}t}}} & 
		{{{\bf{I}}_{{N_{{\rm{RF}}}}}} \!+\! {1 \over {{\sigma ^2}}}{\bf{X}}_{t + 1}^H{{\bf{\Sigma }}_{t}}{{\bf{X}}_{t + 1}}}  \!\!
	\end{array}
	\right]
	\notag 
	\\
	&=I({{{\bf{\bar y}}}_t};{\bf{h}}) + {\log _2}\det \left( 
	{{{\bf{I}}_{{N_{{\rm{RF}}}}}} + {1 \over {{\sigma ^2}}}{\bf{X}}_{t + 
			1}^H{{\bf{\Sigma }}_{t}}{{\bf{X}}_{t + 1}}} \right),
\end{align}
where $(a)$ holds since according to {\bf Lemma \ref{lemma:Orth_W}} and $(b)$ holds by performing matrix triangularization. In particular, ${\bf{\Sigma }}_{t}$ is given by ${{\bf{\Sigma }}_t} = {{\bf{\Sigma }}_{\bf{h}}} - {{\bf{\Sigma }}_{\bf{h}}}{{{\bf{\bar X}}}_t}{\left( {{\bf{\bar X}}_t^H{{\bf{\Sigma }}_{\bf{h}}}{{{\bf{\bar X}}}_t} + {\sigma ^2}{{\bf{I}}_{{N_{{\rm{RF}}}}t}}} \right)^{ - 1}}{\bf{\bar X}}_t^H{{\bf{\Sigma }}_{\bf{h}}}
$, which completes the proof.

%

\begin{figure*}[!b]
	\hrulefill	
	\begin{align}\label{eq:f_v_W_t+1}
		\notag
		f\left( {{{\bf{v}}_{t + 1}},{{\bf{W}}_{t + 1}}} \right)  & 
		\stackrel{(a)}{=} {\log _2}\det \left( {{{\bf{I}}_{{N_{{\rm{RF}}}}}} + 
			{1 \over {{\sigma ^2}}}\sum\limits_{{n_{\rm{T}}} = 1}^{{N_{\rm{T}}}} 
			{\sum\limits_{{n_{\rm{R}}} = 1}^{{N_{\rm{R}}}} {{\lambda 
						_{t,{n_{\rm{T}}},{n_{\rm{R}}}}}\left( {{\bf{v}}_{t + 1}^T \otimes 
						{\bf{W}}_{t + 1}^H} \right)\left( {{{\bf{a}}_{{n_{\rm{T}}}}} \otimes 
						{{\bf{b}}_{{n_{\rm{R}}}}}} \right)\left( {{\bf{a}}_{{n_{\rm{T}}}}^H 
						\otimes {\bf{b}}_{{n_{\rm{R}}}}^H} \right)\left( {{\bf{v}}_{t + 1}^* 
						\otimes {{\bf{W}}_{t + 1}}} \right)} } } \right) \\
		& \stackrel{(b)}{=} {\log _2}\det \left( {{{\bf{I}}_{{N_{{\rm{RF}}}}}} + {1 \over {{\sigma ^2}}}\sum\limits_{{n_{\rm{T}}} = 1}^{{N_{\rm{T}}}} {\sum\limits_{{n_{\rm{R}}} = 1}^{{N_{\rm{R}}}} {{\lambda _{t,{n_{\rm{T}}},{n_{\rm{R}}}}}{{\left| {{\bf{a}}_{{n_{\rm{T}}}}^H{\bf{v}}_{t + 1}^*} \right|}^2}{\bf{W}}_{t + 1}^H{{\bf{b}}_{{n_{\rm{R}}}}}{\bf{b}}_{{n_{\rm{R}}}}^H{{\bf{W}}_{t + 1}}} } } \right)
	\end{align}
\end{figure*}

\section{Proof of Lemma 
	\ref{lemma:Sigma_t_lambda}}\label{appendix:Sigma_t_lambda}
The key idea of the proof is to rewrite the ${{\bf{\bar X}}}_t$-related terms 
in (\ref{eq:Sigma_t}) as ${{\bf{\Sigma }}_{\bf{h}}}{{{\bf{\bar X}}}_t} = 
{{\bf{\Sigma }}_{\bf{h}}}\left[ {{{{\bf{\bar X}}}_{t - 1}},{{\bf{X}}_t}} 
\right]$ and
\begin{align}
	{\bf{\bar X}}_t^H{{\bf{\Sigma }}_{\bf{h}}}{{{\bf{\bar X}}}_t} = \left[ 
	\begin{matrix}
		{{\bf{\bar X}}_{t - 1}^H{{\bf{\Sigma }}_{\bf{h}}}{{{\bf{\bar X}}}_{t - 
					1}}} & {{\bf{\bar X}}_{t - 1}^H{{\bf{\Sigma 
				}}_{\bf{h}}}{{\bf{X}}_t}}  
		\cr 
		{{\bf{X}}_t^H{{\bf{\Sigma }}_{\bf{h}}}{{{\bf{\bar X}}}_{t - 1}}} & 
		{{\bf{X}}_t^H{{\bf{\Sigma }}_{\bf{h}}}{{\bf{X}}_t}}  \cr 
	\end{matrix} \right].
\end{align}
Then, using the Schur's matrix inversion formula to expand the term $\left( 
{{\bf{\bar X}}_t^H{{\bf{\Sigma }}_{\bf{h}}}{{{\bf{\bar X}}}_t} + {\sigma 
		^2}{{\bf{I}}_{{N_{{\rm{RF}}}}t}}} \right)^{ - 1}$ in 
(\ref{eq:Sigma_t}), the following recursion formula of  can be 
obtained:
\begin{align}\label{eq:Sigma_t_recursion}
	{{\bf{\Sigma }}_{t+1}} = {{\bf{\Sigma }}_{t}} - {{\bf{\Sigma 
		}}_{t}}{{\bf{X}}_{t+1}}{\left( {{\bf{X}}_{t+1}^H{{\bf{\Sigma 
				}}_{t}}{{\bf{X}}_{t+1}} + 
			{\sigma ^2}{{\bf{I}}_{{N_{{\rm{RF}}}}}}} \right)^{ - 
			1}}{\bf{X}}_{t+1}^H{{\bf{\Sigma }}_{t}},
\end{align}

When ${{\bf{X}}_{t+1}} = \sqrt{P}{{\bf{U}}_{t}}\left( {:,[1, \cdots 
	,{N_{{\rm{RF}}}}]} \right)$, we have ${{\bf{\Sigma 
	}}_{t}}{{\bf{X}}_{t+1}} 
= 
{\bf{X}}_{t+1}{\rm{diag}}\left( {{\lambda _1}\left( {{{\bf{\Sigma }}_{t}}} 
	\right), \cdots ,{\lambda _{{N_{{\rm{RF}}}}}}\left( {{{\bf{\Sigma }}_{t 
	}}} 
	\right)} \right)$ and ${\bf{X}}_{t+1}^H{{\bf{\Sigma }}_{t 
}}{{\bf{X}}_{t+1}} 
= 
P{\rm{diag}}\left( {{\lambda _1}\left( {{{\bf{\Sigma }}_{t}}} \right), 
	\cdots ,{\lambda _{{N_{{\rm{RF}}}}}}\left( {{{\bf{\Sigma }}_{t}}} 
	\right)} 
\right)$. Thus, the following equality holds:
\begin{align}\label{eq:Sigma_t_U_t-1_X_t}
	\notag
	{{\bf{\Sigma }}_{t+1}} =& {{\bf{U}}_{t}}{{\bf{\Lambda }}_{t}}{\bf{U}}_{t}^H 
	- 
	{{\bf{X}}_{t+1}}{\rm{diag}}{\bigg (} {{\lambda _1^2\left( {{{\bf{\Sigma 
					}}_{t}}} \right)} \over {P{\lambda 
				_1}\left( {{{\bf{\Sigma 
					}}_{t}}} \right) + 
			{\sigma ^2}}},\cdots,\\ &{{\lambda _{{N_{{\rm{RF}}}}}^2( 
			{{{\bf{\Sigma }}_{t}}} )} \over {P{\lambda 
				_{{N_{{\rm{RF}}}}}}\left( 
			{{{\bf{\Sigma }}_{t}}} 
			\right) + {\sigma ^2}}} {\bigg )}{\bf{X}}_{t+1}^H.
\end{align}
Given that ${{\bf{X}}_{t+1}}{\rm{diag}}( {{\lambda _1^2( {{{\bf{\Sigma }}_{t}}} 
		)} \over {P{\lambda _1}( {{{\bf{\Sigma }}_{t}}} ) + {\sigma 
			^2}}}, 
\cdots 
,{{\lambda _{{N_{{\rm{RF}}}}}^2( {{{\bf{\Sigma }}_{t}}} )} \over 
	{P{\lambda 
			_{{N_{{\rm{RF}}}}}}( {{{\bf{\Sigma }}_{t}}} ) + 
		{\sigma ^2}}} 
){\bf{X}}_{t+1}^H 
= {{\bf{U}}_{t}}{\rm{diag}}( {P{\lambda _1^2( {{{\bf{\Sigma }}_{t}}} 
		)} 
	\over {P{\lambda _1}\left( {{{\bf{\Sigma }}_{t}}} \right) + {\sigma 
			^2}}}, 
\cdots ,{P{\lambda _{{N_{{\rm{RF}}}}}^2( {{{\bf{\Sigma }}_{t}}} )} \over 
	{P{\lambda _{{N_{{\rm{RF}}}}}}( {{{\bf{\Sigma }}_{t}}} ) + {\sigma 
			^2}}},\!\!\underbrace {0, \cdots ,0}_{\small N_{\rm 
		R}N_{\rm 
		T}-N_{\rm 
		RF}}\!\!){\bf{U}}_{t}^H$ and ${{\bf{\Sigma }}_{t}} = 
{{\bf{U}}_{t}}{{\bf{\Lambda }}_{t}}{\bf{U}}_{t}^H$, the equality 
in 
(\ref{eq:Sigma_t_U_t}) can be derived from (\ref{eq:Sigma_t_U_t-1_X_t}), which 
completes the proof.

\section{Proof of Corollary 
	\ref{corollary:Sigma_h_aT_bR}}\label{appendix:Sigma_h_aT_bR}
According to {\bf Lemma \ref{lemma:Sigma_decomp}} and equality 
$\left( {{\bf{AB}}{{\bf{A}}^H}} \right) \otimes \left( {{\bf{CD}}{{\bf{C}}^H}} 
\right) = \left( {{\bf{A}} \otimes {\bf{C}}} \right)\left( {{\bf{B}} \otimes 
	{\bf{D}}} \right)\left( {{{\bf{A}}^H} \otimes {{\bf{C}}^H}} \right)$, the 
kernel ${{\bf{\Sigma}}_{\bf h}}$ can be decomposed as 
\begin{align}\label{appendix:sigma_h}
	\notag
	{{\bf{\Sigma }}_{\bf{h}}} = &  \left({{\bf{U}}_{\rm{T}}} {{\bf{\Lambda 
		}}_{\rm{T}}}  {{\bf{U}}_{\rm{T}}}^H\right) \otimes \left({\bf{U}}_{\rm{T}} 
	{\bf \Lambda 
	}_{\rm{T}}  {{\bf{U}}_{\rm{T}}}^H\right)
	\\ \notag
	= & \underbrace {\left( {{{\bf{U}}_{\rm{T}}} \otimes 
			{{\bf{U}}_{\rm{R}}}} \right)}_{
		\mb{U}_0}\underbrace {\left( 
		{{{\bf{\Lambda }}_{\rm{T}}} \otimes {{\bf{\Lambda }}_{\rm{R}}}} 
		\right)}_{\text{Eigenvalue matrix}}\left( {{\bf{U}}_{\rm{T}}^H \otimes 
		{\bf{U}}_{\rm{R}}^H} \right) \\
	= &  \sum\limits_{{n_{\rm{T}}} = 1}^{{N_{\rm{T}}}} 
	{\sum\limits_{{n_{\rm{R}}} = 1}^{{N_{\rm{R}}}} {{\alpha 
				_{{n_{\rm{T}}}}}{\beta _{{n_{\rm{R}}}}}\left( {{{\bf{a}}_{{n_{\rm{T}}}}} 
				\otimes {{\bf{b}}_{{n_{\rm{R}}}}}} \right){{\left( 
					{{{\bf{a}}^H_{{n_{\rm{T}}}}} \otimes {{\bf{b}}^H_{{n_{\rm{R}}}}}} 
					\right)}}} },
\end{align}
Based on \eqref{appendix:sigma_h}, one can verify without difficulty that 
(\ref{eq:Sigma_h_T_R}) is exactly the eigenvalue decomposition of ${{\bf{\Sigma 
	}}_{\bf h}}$, which completes the proof.

\section{Proof of Lemma \ref{lemma:eigen_selection}}\label{appendix:eigen_selection}

Given the new constraints ${{\bf{v}}_{t+1}}\in\left\{\sqrt{P} {\bf{a}}^*_{{n_{\rm{T}}}} \right\}_{{n_{\rm{T}}} = 1}^{{N_{\rm{T}}}}$ and ${\bf w}_{t+1,k}\in\left\{ {\bf{b}}_{{n_{\rm{R}}}} \right\}_{{n_{\rm{R}}} = 1}^{{N_{\rm{R}}}}$ for all $k\in\{1,\cdots,N_{\rm RF}\}$, problem (\ref{eq:problem_t+1}) can be reorganized as
\begin{align}\label{eq:probelm_appendix_derive}
	\notag
	\mathop {\max }\limits_{ {\bf{v}}_{t+1}, {\bf W}_{t+1}}~
	&f\left( {{{\bf{v}}_{t + 1}},{{\bf{W}}_{t + 1}}} \right) \\
	\notag
	{\rm s.t.}~&{{\bf{v}}_{t+1}}\in\left\{\sqrt{P} {\bf{a}}^*_{{n_{\rm{T}}}} \right\}_{{n_{\rm{T}}} = 1}^{{N_{\rm{T}}}}, 
	\\
	\notag
	&{\bf w}_{t+1,k}\in\left\{ {\bf{b}}_{{n_{\rm{R}}}} \right\}_{{n_{\rm{R}}} = 1}^{{N_{\rm{R}}}}, \forall k\in\{1,\cdots,N_{\rm RF}\},
	\\
	&{\bf w}_{t+1,k} \ne {\bf w}_{t+1,k'}, \forall k\ne k',
\end{align}
where the objective function is given in (\ref{eq:f_v_W_t+1}), in which $(a)$ holds according to the definition in (\ref{eq:Sigma_t_definition}) and $(b)$ holds by utilizing the property to the property that $\left( {{\bf{A}} \otimes {\bf{B}}} \right) \left( {{\bf{C}} \otimes {\bf{D}}} \right) = \left( {{\bf{AC}}} \right) \otimes \left( {{\bf{BD}}} \right)$. Note that, the constraint ${\bf w}_{t+1,k} \ne {\bf w}_{t+1,k'}$ for all $k\ne k'$ in (\ref{eq:probelm_appendix_derive}) ensures the orthogonality of ${{\bf{W}}_{t + 1}}$. Observing (\ref{eq:probelm_appendix_derive}), one can find that our goal becomes finding optimal indexes $n_{\rm T}$ and $\left\{ {{n_{{\rm{R}},k}}} \right\}_{k = 1}^{{N_{{\rm{RF}}}}}$ that maximize the MI increment $f\left( {{{\bf{v}}_{t + 1}},{{\bf{W}}_{t + 1}}} \right)$. Assuming that the optimal indexes are expressed by $n_{\rm T}^{\rm opt}$ and $\{ {{n^{\rm opt}_{{\rm{R}},k}}} \}_{k = 1}^{{N_{{\rm{RF}}}}}$, the optimal precoder and the optimal combiner are 
\begin{align}
	{\bf{v}}_{t + 1}^ {\rm opt}  = \sqrt{P}{\bf{a}}_{n_{\rm{T}}^ {\rm opt} }^* ~\text{and}~ 
	{\bf{W}}_{t + 1}^ {\rm opt}  = \left[{{{\bf{b}}_{n_{{\rm{R}},1}^ {\rm opt} }}, \cdots 
		,{{\bf{b}}_{n_{{\rm{R}},{N_{{\rm{RF}}}}}^ {\rm opt} }}}\right],
\end{align}
respectively. Then, we have
\begin{subequations}\label{eq:av_bW}
	\begin{align}
		{ {\bf{a}}_{n_{\rm{T}}}^H({\bf{v}}^{\rm opt}_{t + 1})^* } &= \left\{ {\begin{matrix}
				\sqrt{P}, & n_{\rm T} = n^{\rm opt}_{\rm T}  \cr 
				0, & {{\text{else}}}  \cr 
		\end{matrix} } \right.,
		\\
		{{\bf{b}}_{{n_{\rm{R}}}}^H{{\bf{W}}^{\rm opt}_{t + 1}}} &= \left\{ {\begin{matrix}
				{\bf e}^T_{n_{R}}, & n_{\rm R} \in \{ {{n^{\rm opt}_{{\rm{R}},k}}} \}_{k = 1}^{{N_{{\rm{RF}}}}} \cr 
				{\bf 0}^T_{N_{\rm RF}}, & {{\text{else}}}  \cr 
		\end{matrix} } \right.,
	\end{align}
\end{subequations}
where ${\bf e}_{n_{R}}$ denotes an $N_{\rm RF}$-dimensional vector whose $n_{R}$-th entry is one and the other entries are zero. By substituting (\ref{eq:av_bW}) into (\ref{eq:probelm_appendix_derive}), the optimal MI increment $f\left( {{{\bf{v}}^{\rm opt}_{t + 1}},{{\bf{W}}^{\rm opt}_{t + 1}}} \right)$ can be expressed by
\begin{align}
	\notag
	&f\left( {{{\bf{v}}^{\rm opt}_{t + 1}},{{\bf{W}}^{\rm opt}_{t + 1}}} \right)
	\\ 
	\notag
	= & \log_2 \det \left( {{{\bf{I}}_{{N_{{\rm{RF}}}}}} + {P \over {{\sigma 
					^2}}}\sum\limits_{{n_{\rm{R}}} = 1}^{{N_{\rm{R}}}} {{\lambda 
				_{t,{n^{\rm opt}_{\rm{T}}},{n_{\rm{R}}}}} ({\bf{W}}^{\rm opt}_{t + 
				1})^H {{\bf{b}}_{{n_{\rm{R}}}}}{\bf{b}}_{{n_{\rm{R}}}}^H{{\bf{W}}^{\rm opt}_{t + 1}}} } 
	\right) 
	\\ 
	\notag
	= &
	\log_2\det \left( {{{\bf{I}}_{{N_{{\rm{RF}}}}}} + {P \over {{\sigma 
					^2}}}{\rm{diag}}\left( {{\lambda 
				_{t,{n^{\rm opt}_{\rm{T}}},{n^{\rm opt}_{{\rm{R}},1}}}}, \cdots ,{\lambda 
				_{t,{n^{\rm opt}_{\rm{T}}},{n^{\rm opt}_{{\rm{R}},{N_{{\rm{RF}}}}}}}}} \right)} \right)
	\\ = &
	\sum\limits_{k = 1}^{{N_{{\rm{RF}}}}} {{{\log }_2}\left( {1 + {{{P\lambda 
						_{t,{n^{\rm opt}_{\rm{T}}},{n^{\rm opt}_{{\rm{R}},k}}}}} \over {{\sigma ^2}}}} \right)},
\end{align}
which only relies on the eigenvalues of ${\bf \Sigma}_t$. In this context, the problem becomes finding $n_{\rm T}$ and $\{ {{n_{{\rm{R}},k}}} \}_{k = 1}^{{N_{{\rm{RF}}}}}$ that maximize $f\left( {{{\bf{v}}_{t + 1}},{{\bf{W}}_{t + 1}}} \right)$, as formulated in (\ref{eq:problem_t+1_reorg}). This completes the proof.

\footnotesize
\bibliographystyle{IEEEtran}
	
\begin{spacing}{0.98}
\bibliography{IEEEabrv,reference}	
\end{spacing}
	
	
\end{document}